\documentclass[prd,aps,floats,floatfix,eqsecnum,nofootinbib]{revtex4}
\usepackage{verbatim,graphicx,amssymb,amsbsy,bm,amsmath,rotating,epsfig}
\usepackage{psfrag}
\usepackage{hyperref}
\usepackage{fontenc}
\usepackage[latin1]{inputenc}
\usepackage{pstricks,pst-node,pst-text,pst-3d}
\usepackage{textcomp}
\newcommand{\be}{\begin{equation}}
\newcommand{\ee}{\end{equation}}
\newcommand{\bea}{\begin{eqnarray}}
\newcommand{\eea}{\end{eqnarray}}

\begin{document}
\title{The Dark Matter distribution function and Halo Thermalization from the Eddington equation in Galaxies}
\author{\bf H. J. de Vega $^{(a,b)}$}
\email{devega@lpthe.jussieu.fr} 
\author{\bf N. G. Sanchez $^{(b)}$}
\email{Norma.Sanchez@obspm.fr} 
\affiliation{$^{(a)}$ LPTHE, Universit\'e Pierre et Marie Curie (Paris VI),
Sorbonne Universit\'es, 
Laboratoire Associ\'e au CNRS UMR 7589, Tour 24, 5\`eme. \'etage, 
Boite 126, 4, Place Jussieu, 75252 Paris, Cedex 05, France. \\
$^{(b)}$ LERMA, CNRS UMR 8112, Observatoire de Paris PSL, Sorbonne Universit\'es, UPMC Univ. Paris VI. 
\\ 61, Avenue de l'Observatoire, 75014 Paris, France.}
\date{\today}
\begin{abstract}
We find the distribution function $ f(E) $ for dark matter (DM) halos in galaxies 
and the corresponding equation of state from the (empirical) DM density
profiles derived from observations. We solve for DM in galaxies the analogous of the
Eddington equation originally used for the gas of stars in globular clusters. The observed 
density profiles are a good realistic starting point and the distribution functions derived 
from them are realistic. We do not make any assumption about 
the DM nature, the methods developed here apply to any DM kind, though
all results are consistent with Warm DM (WDM). With these methods we find:
(i) Cored density profiles behaving quadratically for small distances 
$ \rho(r) \buildrel_{r \to 0}\over=\rho(0) - K \; r^2 $ produce distribution functions
which are finite and positive at the halo center while cusped density profiles always produce
divergent distribution functions at the center.
(ii) Cored density profiles produce approximate thermal Boltzmann distribution functions for $ r \lesssim 3 \; r_h $
where $ r_h $ is the halo radius. (iii) Analytic expressions for the dispersion velocity
and the pressure are derived yielding at each halo point an ideal DM gas equation of 
state with local temperature $ T(r) \equiv m \; v^2(r)/3 . \;
 T(r) $ turns to be constant in the same region where the distribution function is thermal and exhibits the same temperature
within the percent. The self-gravitating DM gas can thermalize despite being collisionless because
it is an ergodic system. (iv) The DM halo can be consistently considered at local thermal equilibrium with:
(a) a constant temperature  $ T(r) = T_0 $ for $ r \lesssim 3 \; r_h $, (b) a space dependent 
temperature $ T(r) $ for $ 3 \; r_h < r \lesssim R_{virial} $, which slowly decreases with $ r $.
That is, the DM halo is realistically a collisionless self-gravitating thermal gas for $ r \lesssim R_{virial} $. 
(v) $ T(r) $ outside the halo radius nicely follows the decrease of the circular velocity squared. 
\end{abstract}
\pacs{95.35.+d, 98.80.-k, 98.80.Cq}
\keywords{Dark Matter, Galaxy structure, Galaxy Density Profiles}
\maketitle
\tableofcontents

\section{INTRODUCTION}

Dark matter (DM) is the main component of galaxies: the fraction of DM over the total
galaxy mass goes from 95\% for large diluted galaxies till 99.99\% for dwarf compact galaxies.
Therefore, DM alone should explain the main structure of galaxies. Baryons should only give
corrections to the pure DM results. Astronomical observations show that the DM galaxy density 
profiles are {\bf cored} till scales below the kpc \cite{obs,gil,wp}.

\medskip

Galaxies first form by  collapse of primordial density fluctuations through processes far from
stationary \cite{bin}. In this early evolution two well known mechanisms come into play: phase mixing and 
violent relaxation. D. Lynden-Bell \cite{lyn}, S. Tremaine  et al.  \cite{stre} and
D. Merrit et al.  \cite{mer} proposed physically-motivated distribution
functions to capture the physics of violent relaxation. 
N-body numerical simulations are able to follow quantitatively the galaxy all along till today.
Cold dark matter (CDM) particles heavier than a GeV spectacularly succeed to reproduce the observations for large
scales beyond the Mpc. CDM and warm dark matter (WDM) yield identical results 
for scales beyond the Mpc.

\medskip

In this paper we do not investigate the galaxy evolution. 
We consider DM dominated galaxies in their late stages of structure formation when 
they are relaxing to a stationary situation, at least inside the virial radius.

{\vskip 0.1cm} 

This is a realistic situation since the free-fall (Jeans) time $ t_{ff} $ for galaxies
is much shorter than the age of galaxies:
$$
t_{ff} = \frac1{\sqrt{G \; \rho_0}} = 1.49 \; 10^7 \; 
\sqrt{\frac{M_\odot}{\rho_0 \; {\rm pc}^3}} \; {\rm yr} \; .
$$
The observed central densities of galaxies yield free-fall times in the range
from 15 million years for ultra-compact galaxies till 330 million years for
large diluted spiral galaxies. These free-fall (or collapse) times are small compared
with the age of galaxies running in billions of years.

{\vskip 0.1cm} 

Hence, we can consider the DM described by a time independent and non--relativistic 
energy distribution function $ f(E) $, where $ E = p^2/(2m) - \mu $
is the single--particle energy, $ m $ is the mass of the DM particle and
$ \mu $ is the chemical potential \citep{newas,astro}.

{\vskip 0.1cm} 

For collisionless self-gravitating systems as the DM galaxy halo, the DM chemical potential
$ \mu(r) $ is proportional to the gravitational potential $ \phi(r) $,
\be\label{potq}
\mu(r) =  \mu_0 - m \; \phi(r)  \; , 
\ee
$ \mu_0 $ being a constant, and $ \mu(r) $
obeys the {\bf self-consistent} and {\bf nonlinear} Poisson equation
\be\label{pois7}
\nabla^2 \mu(r) = -4 \; \pi \; g \; m \; \rho(r)
=-4 \; \pi \; g \; G \; m^2 \; 
\int \frac{d^3p}{(2 \, \pi \; \hbar)^3} \; f\left(\frac{p^2}{2 \, m}-\mu(r)\right) \; .
\ee
Here $ G $ is Newton's gravitational constant, $ g $ is the number of internal degrees of freedom 
of the DM particle, $ p $ is the DM particle momentum and $ f(E) $ is the DM
energy distribution function (DF). 

{\vskip 0.1cm} 

In this paper spherical symmetry is considered for simplicity,
It is clear that DM halos are not perfectly
spherical but describing them as spherically symmetric is a first
approximation that can be improved. 

{\vskip 0.1cm} 

Our spherically symmetric treatment captures the essential features
of the gravitational structure and agree with the observations.
As discussed in the Introduction, 
it is known from observations that the angular momentum of DM halos is small and therefore
DM halos can be considered spherically symmetric \cite{ton,peebu}. 
Hence, their phase--space distribution function can be considered only a function of the energy. 
This approach can be extended to obtain distribution functions depending on other parameters as the 
angular momentum. Ansatze for phase--space distribution functions which also depend on the 
angular momentum have been considered in \cite{bin,pad,bin2,ngu}.

{\vskip 0.1cm} 

It is safe to ignore the presence of a central black holes (BH) to 
describe the DM halos structure. BH are at the galaxy center 
and the abundant literature of observed cored density profiles
describe the profiles without central BH.   

\medskip

In refs. \citep{newas,astro,dVSS2,nosoct} we proposed and developed the Thomas-Fermi approach to galaxy structure 
for self-gravitating fermionic warm DM (WDM). Given the distribution function $ f(E) $, this approach
allows to obtain self-consistently the potential $ \mu(r) $ from eq.(\ref{pois7}), as well 
as the density profiles, velocity dispersion and the equation of state.

\medskip

In the present paper our aim is to gain knowledge on the central object describing halos:
the DM distribution function $ f(E) $. For this purpose, instead to compute the density profiles 
and galaxy relevant structural magnitudes from a given distribution function
$ f(E) $, we {\bf find} here the distribution function $ f(E) $, 
and the corresponding equation of state, from the DM halo density
profiles derived from observations $ \rho(r) $. Namely, we determine and solve for DM in galaxies the Eddington equation originally used for 
the gas of stars in globular clusters. The DM profiles derived from observations
are good realistic starting points
and the distribution functions derived from them are realistic ones.
 
{\vskip 0.1cm} 

Observations, analytic estimations and simulations 
show that the DM angular momentum in galaxies is small \cite{ton,peebu}.
The spin parameter, $ \lambda = J |E|/[G M^{5/2}] $
where $ J, \; E $ and $ M $ are the total angular momentum, energy and mass
of the system, and G in Newton's constant, 
takes values of the order of the percent \cite{ton,peebu}.
In spirals we have a direct proof of this fact from their bottom
up general scenario of formation. In these objects we can
compute from observations the disk angular momentum, if
the angular momentum per unit mass is
conserved during the process of disk formation, the values
found imply that DM halos are not dominated by rotation \cite{ton}.
In ref. \cite{newas} we estimated the angular momentum
effect for compact galaxies and this yields small corrections.
Therefore, the spherical symmetry is a good approximation 
to start with. Indeed, our results confirm the consistency of such assumption.
The methods of this paper can be generalized to anisotropic distributions in 
which case the distribution function will depend on the angular momentum too 
\cite{bin,pad} although we will not consider it in this paper. 

{\vskip 0.1cm} 

No assumption is made here about the mass of the DM particle, the formalism developed in this paper
applies to any kind of DM. In any case, all our results are consistent with a DM particle mass in 
the keV scale (WDM).

{\vskip 0.1cm} 

We do not assume any equation of state but we derive and compute it from the distribution function
with the general kinetic theory expression \cite{newas,astro,bdvs}
$$
P(r) = \frac13 \; v^2(r) \; \rho(r) \; ,
$$
where $ P(r) $ is the pressure.

{\vskip 0.1cm} 

The variation of the velocity dispersion with galactocentric radii $ r $ 
allows us here to obtain the equation of state of the DM which is obviously
a result of high interest.

\medskip

The short distance behaviour of the density profile determines through the solution
of the Eddington equation the behaviour of the distribution function $ f(E) $ for the lowest possible energy $ E $.
We explicitly compute $ f(E) $ for the lowest possible energy both for cored and cusped density profiles.
For cored profiles only the quadratic behaviour for small distances:
\be\label{ufaI}
\rho(r) \buildrel_{r \to 0}\over=\rho(0) - K \; r^2 \; ,
\ee
where $ K $ is a positive constant, provides a physical meaningful distribution function near the center,
namely, a positive and bounded distribution function $ f(E) $ for the lowest energy $ E $.
It was noticed in ref. \cite{posti} that cored profiles can give bounded distribution functions (DF) at the center.
We provide here for the first time the precise condition eq.(\ref{ufaI}) to have a bounded DF at the center for cored
profiles.

{\vskip 0.1cm} 

The Eddington equation has been applied to to derive the DF from the density profile in several articles
\cite{huq,evan,wid,loma,posti}. 
For cusped profiles the distribution function always diverges near the halo center. This was noticed in ref. \cite{wid,loma,posti}
for specific cusped models (Hernquist, Jaffe, NFW).

{\vskip 0.1cm} 

The distribution function for spin-$1/2$ fermions must be everywhere smaller than two
because of the Pauli principle. Cored density profiles with the behaviour eq.(\ref{ufaI})
satisfy the Pauli bound while cusped profiles always violate the Pauli principle 
near the halo center.

\medskip

We explicitly compute in this paper the phase--space distribution function and the equation
of state for the family of $\alpha$-density profiles
\be\label{rhoalfa}
 \rho(r) = \frac{\rho_0}{\left[1+\left(\displaystyle\frac{r}{r_h}\right)^2\right]^\alpha} 
\quad , \quad 1 \leq \alpha \leq 2.5 \; .
\ee
This cored density profile is a generalization of the pseudo-thermal profile and 
with $ \alpha \sim 1.5 $, it is perfectly appropriate to fit galaxy observations.

{\vskip 0.1cm} 

For $ \alpha \sim 1.5 $, we find that inside the halo radius $ r < r_h $, the ratio 
$$ 
\frac{P(r)}{\rho(r)} = \frac13 \; v^2(r)
$$ 
turns to be {\bf approximately constant} (independent of $ r $). Therefore, the local temperature
\be\label{tcirI}
T(r) \equiv \frac13 \; m \;  v^2(r)
\ee
is {\bf approximately constant} $ T(r) \simeq T_0 $ inside the halo radius
as shown by figs. \ref{psobrerho} and \ref{tef}. Moreover, this implies that the dark matter obeys 
locally, at each point of the halo, the ideal gas equation of state
\be\label{TqI}
P(r) = \frac{T(r)}{m} \; \rho(r)  \; .
\ee
Furthermore, the distribution function turns to be Boltzmann-like in the region 
$ 0 \leq r \lesssim  3 \; r_h $ with the same temperature $ T_0 $ that in 
the equation of state eq.(\ref{TqI}) showing the consistency of the thermalization.

In summary, on the basis of galaxy observations, mainly of the galaxy density profiles,
and without any assumption about 
the DM particle mass, we find that the DM in the galaxy halos is
approximately thermalized for $ r \lesssim 3 \; r_h $. That is, the DM is a collisionless
self-gravitating thermal gas in a region extending beyond the halo 
radius although much smaller than the virial radius $ R_{virial} $.

{\vskip 0.2cm} 

In the region $ 3 \; r_h < r \lesssim R_{virial} $, DM is a virialized collisionless self-gravitating gas which
can be consistently considered at local thermal equilibrium with a space dependent 
temperature $ T(r) $ that slowly decreases with the distance to the center $ r $ as shown in fig. \ref{tef}.

{\vskip 0.1cm}

We introduce the circular temperature $ T_c(r) $ associated to the circular velocity in the same way
that the temperature  $ T(r) $ is associated to the velocity through eq.(\ref{tcirI}). 
We find that the local temperature profile $ T(r) $ nicely
{\bf follows} the decrease of the circular temperature  $ T_c(r) $ in the region $ r \gtrsim  r_h $
as shown in fig. \ref{tef}.

{\vskip 0.2cm} 

Realistic (empirical) density profiles show that the self-gravitating DM in the inner halo region
$ r \lesssim r_h $ is approximately thermalized despite the DM being collisionless.

{\vskip 0.2cm} 

The collisionless self-gravitating gas is an isolated system which is not integrable.
Therefore, it is an ergodic system that can thermalize \cite{gall}.
Namely, the particle trajectories explore
ergodically the constant energy manifold in phase-space, covering it uniformly
according to precisely the microcanonical measure and yielding to a thermal situation \cite{gall}.

{\vskip 0.2cm} 

Physically, these phenomena are clearly understood because in the inner halo region
$ r \lesssim r_h $, the density is higher than beyond the halo radius.
The gravitational interaction in the inner region is strong enough and
thermalizes the self-gravitating gas of DM particles while beyond the halo radius
the particles are too dilute to thermalize, namely, although they are virialized,
they had not enough time to accomplish thermalization. Notice that virialization always starts 
before than thermalization. 

{\vskip 0.1cm} 

In the process of thermalization there is an energy transfer flow of potential energy into kinetic energy.
Clearly, in the outside halo region $ r \gtrsim  3 \; r_h $ we find that the kinetic energy is lower
than in the inside the region $ r < r_h $ where thermalization is already achieved.
Therefore, the local temperature $ T(r) $ in the outside halo region $ r \gtrsim  3 \; r_h $
is lower than the temperature $ T_0 $ in the internal region $ r < r_h $ where thermalization is achieved.

{\vskip 0.1cm} 

The dynamical study of the thermalization mechanism for long-range self-gravitating forces
is an interesting important issue beyond the scope of this paper. In this respect, extending
to long-range forces the work ref. \cite{DdV} on thermalization would be appropriate.

{\vskip 0.1cm} 
 
The treatment presented in this paper applies to dilute large galaxies which are in a classical
physics regime for halo masses $ M_h > 10^6 \; M_\odot $.

{\vskip 0.1cm} 
 
For smaller (dwarf) galaxies there is not yet available information on density profiles
from observations. Knowing the density profiles for dwarf galaxies
will allow to apply the framework provided in this paper to find the phase--space distribution
function $ f(E) $ and the velocity dispersion $ v^2(r) $ for dwarf galaxies.

{\vskip 0.1cm} 

On the other hand, the Thomas-Fermi approach to galaxy structure
applies to all types of galaxies and allows to determine
theoretically all physical magnitudes for them \cite{newas}-\cite{astro}.
It must be noticed that in the classical regime, for halo masses $ M_h > 10^6 \; M_\odot $, the
galaxy equation of state computed in the Thomas-Fermi approach yields
the same results as found here from the empirical $\alpha $-profiles
and the Eddington equation, namely the Boltzmann self-gravitating gas,
showing the robustness of these results.

{\vskip 0.1cm} 
 
For dwarf galaxies, the equation of state derived in the Thomas-Fermi theory
exhibits deviations from the Boltzmann ideal gas equation of state
due to macroscopic quantum effects well accounted by the 
Thomas-Fermi approach \cite{newas}-\cite{astro}
and which reflect the quantum fermion state near degeneration characteristic
of dwarf compact galaxies.

{\vskip 0.1cm} 
 
This paper deals with  the classical regime because
the phase-space DF function of $ p $ AND $ q $ yields a classical description of the collisionless
self-gravitating gas. 

\medskip 

The density profile eq.(\ref{rhoalfa}) for $ \alpha = 2.5 $ is not appropriate to describe 
DM halos but corresponds to the Plummer profile describing the density of stars 
in globular clusters \cite{plu,bin}. We find that the distribution function for the Plummer 
 profile is approximately thermal only for a small range around the center. That is,
stars in globular clusters are approximately thermal
in a narrower region both in energy and coordinates than the DM in galaxy halos.

\medskip 

We provide here an unified framework in which galaxy structure is
obtained from the self-gravitating gas of DM particles in the same way that the  
globular cluster structure is obtained from a gas of stars.

\medskip

This paper is organized as follows. In Section 2 we solve the Eddington equation for Dark Matter in Galaxies,
we express the velocity dispersion and the pressure in terms of the density profile and compute
the distribution function behaviour near the center in cored and cusped profiles. In section 3
we explicitly compute the distribution function for the family of $\alpha $-density profiles 
eq.(\ref{rhoalfa}) showing the approximate thermalization inside the halo radius for $ \alpha \sim 1.5 $. 
In section 4 we compute the DM equation of state and confirm the approximate thermalization 
inside the halo radius. In section 5 we show that the DM outside the halo radius is in local
thermal equilibrium with a local temperature slowly decreasing with $ r $.

\section{The Eddington equation for Dark Matter in Galaxies}

DM is the main component of galaxies: the DM fraction over the total mass
goes from 95\% for large diluted galaxies \cite{pers} till 99.99\% for dwarf compact
galaxies. Astronomical observations show that DM galaxy density profiles
are cored till scales below the kpc \cite{woo}. 
At these scales, baryons cannot transform the DM cores of the density profiles into cusps.
The DM cored profiles obtained from galaxy observations are the starting 
point in the present Eddington-like approach,
the profiles here are not assumptions but they are
a truly realistic starting basis provided by the observations \cite{obs,gil,wp,woo}.

{\vskip 0.1cm} 

The self-gravity of the baryonic material is negligible while
baryons are immersed in a DM halo potential well.
Baryons trace the DM potential well playing the role of test
particles to measure the local DM density.

\medskip

The mass density $ \rho(r) $ is expressed as a function of the chemical potential $ \mu(r) $ through the
standard integral of the DM phase--space distribution function over the momentum
\be \label{den}
  \rho(r) = \frac{g \, m}{2 \, \pi^2 \, \hbar^3} \int_0^{\infty} dp\;p^2 
  \; f\left[\displaystyle \frac{p^2}{2m}-\mu(r)\right] \; , 
\ee
where $ g $ is the number of internal degrees of freedom of the DM particle,
with $ g = 1 $ for Majorana fermions and $ g = 2 $ for Dirac fermions. For
definiteness, we will take $g=2$ in the sequel.

{\vskip 0.1cm} 

The Poisson equation for the gravitational potential $ \phi(r) $ takes the self-consistent form eq.(\ref{pois7}),
\be \label{pois}
  \frac{d^2 \mu}{dr^2} + \frac2{r} \; \frac{d \mu}{dr} = - 4\pi \, G \, m \, \rho(r) =
- \frac{4 \; G \; m^2}{\pi \; \hbar^3} \int_0^{\infty} dp\;p^2 
  \; f\left[\displaystyle \frac{p^2}{2m}-\mu(r)\right]\; , 
\ee
where $ G $ is Newton's constant and we used eq.(\ref{potq}).

\medskip

It is useful to introduce dimensionless variables $ q , \; \nu(q) $ 
\be\label{varsd}
 r = r_h \; q \quad , \quad \mu(r) =  T_0 \;  \nu(q) \quad . 
\ee
Following eq.(\ref{rhoalfa}), we define the core size $ r_h $ of the halo as
\be\label{onequarter}
  \frac{\rho(r_h)}{\rho(0)} = 2^{-\alpha} \quad .
\ee
$ T_0 $ is the characteristic one--particle energy scale. $ T_0 $ plays
the role of an effective temperature scale and depends on the galaxy mass.

\medskip

We consider the density profile
\be\label{denF}
\rho(r) = \rho_0 \; F\left(\frac{r}{r_h}\right) = \rho_0 \; F(q) \quad , \quad \rho_0 \equiv \rho(0)
\; ,
\ee
where the given function $ F(q) $ takes a bounded value at the origin $ F(0) = 1 $.
That is, we typically consider cored density profiles although our treatment applies for any density profile.

{\vskip 0.1cm} 

Then, in dimensionless variables the Poisson's equation eq.(\ref{pois}) takes the form
\be\label{defb}
\frac{d^2 \nu}{dq^2} + \frac2{q} \; \frac{d \nu}{dq} = - b_0\; F(q) \quad , \quad
b_0\equiv 4 \; \pi \; G \; \rho_0 \; r_h^2 \; \frac{m}{T_0} \; .
\ee
This equation can be solved in closed form with the solution
\be\label{nuexp}
\nu(q) = \nu(0) - b_0\; \int_0^q\left(1 -  \frac{q'}{q} \right) q' \;  F(q') \; dq' 
 \quad , \quad \frac{d\nu}{dq} = - \frac{b_0}{q^2} \; \int_0^q q'^2 \;  F(q') \; dq' \quad  .
\ee

In dimensionless variables, for spherically symmetric distributions, eq.(\ref{den}) becomes
\be\label{den2}
\rho(r) = \frac{\sqrt2}{\pi^2} \; m^\frac52 \; T_0^\frac32 \int_{\nu(\infty)}^{\nu}  d\nu' \; \sqrt{\nu-\nu'} 
\; \; \Psi(-\nu') \quad , \quad \nu' \equiv \nu - \frac{p^2}{2 \, m \;  T_0} 
\quad , \quad \Psi(-\nu) \equiv f(-T_0 \; \nu) \; .
\ee
$ \nu(q) $ takes its minimum value at $ q = \infty $. This is so because $ d\nu(q)/dq < 0 $
as follows from eq.(\ref{nuexp}). Therefore, this minimum value $ \nu(\infty) $ is 
the lower bound of integration in $ \nu' $. 
The chemical potential for galaxies is always negative except for the dwarf compact galaxies 
in the quantum regime \cite{nosoct}. Hence, $ \nu(\infty) $ is always negative except 
in the limiting case of degenerate WDM fermions \cite{nosoct}.

{\vskip 0.1cm} 

Eq.(\ref{den2}) can thus be written as
\be \label{den3}
F(\nu) = \frac{\sqrt2}{\pi^2}\; \frac{m^\frac52 \; T_0^\frac32}{\rho_0} \;
\int_{\nu(\infty)}^\nu  d\nu' \; \sqrt{\nu-\nu'} \; \; \Psi(-\nu') \; .
\ee

Eq.(\ref{den3}) expressing the density profile in terms of the distribution function
is the {\it Abel integral} equation and can be explicitly inverted as \cite{gel}, 
\be\label{psinu}
\Psi(-\nu) = \sqrt2 \; \pi \; \frac{\rho_0}{m^\frac52 \; T_0^\frac32} \;
\int_{\nu(\infty)}^\nu \frac{d\nu'}{\sqrt{\nu-\nu'}} \; \frac{d^2 F}{d\nu'^2} \; ,
\ee
with $ \Psi $ and $ d\Psi/d\nu $ vanishing at infinite distance $ q = \infty $ as boundary condition.

\medskip

Eq.(\ref{psinu}) is the {\it Eddington formula}, originally derived for a gas of stars forming a 
globular cluster \cite{eddi} also valid for collisionless self-gravitating DM halos of galaxies.
Given the density profile $ F(q) $, the distribution function $ \Psi(-\nu) $ follows by quadratures. 
The Eddington formula has been applied to derive the DF from the density profile in
\cite{huq,evan,wid,loma,posti}. 

\medskip

It is useful to re-scale the potential $ \nu(q) $ as
\be\label{defeps}
\nu(q) = \nu(0) + b_0 \; \varepsilon(q) \; ,
\ee
with the coefficient $ b_0 $ given by eq.(\ref{defb}). Then, eq.(\ref{nuexp}) yields
\be \label{eqst}
\varepsilon(q) =-\int_0^q\left(1 -  \frac{q'}{q} \right) q' \;  F(q') \; dq' \quad , \quad
\frac{d\varepsilon}{dq} = - \frac1{q^2} \; \int_0^q q'^2 \;  F(q') \; dq' \quad , \quad \varepsilon(0) = 0 \; .
\ee
$ \varepsilon(q) $ is related to the potential energy $ E(q) $ of a particle at the point
$ q $ in the galaxy by
\be \label{Eq}
E(q) =  E(0) - 4 \; \pi \; G \; \rho_0 \; r_h^2 \; m \; \varepsilon(q) 
\ee
Hence, $ -\varepsilon(q) $ is the particle potential energy in units of $ 4 \; \pi \; G \; \rho_0 \; r_h^2 \; m $
taking as reference the potential energy at the origin. We choose the potential energy at the center to be zero
$  E(0) = 0 $. ???????????

\medskip

Since the profile $ F $ is given explicitly as a function of $ q $, it is
convenient in eq.(\ref{psinu}) to change the integration variable from $ \nu' $ to $ q' $ 
using eqs.(\ref{defeps})-(\ref{eqst}) with the result
\bea\label{solabel}
&& \Psi(q) = \frac1{G^\frac32 \; r_h^3 \; m^4 \; \sqrt{\rho_0}} \; \; {\cal D}(q) \quad ,\quad
{\cal D}(q) \equiv  \frac1{\sqrt{32 \; \pi}}
\int_q^{\infty} \frac{ {\cal J}(q') \;  dq'}{\sqrt{\varepsilon(q)-\varepsilon(q')}} \;
\\ \cr 
&&  {\rm where} \qquad {\cal J}(q) \equiv 
\frac1{\left(-\displaystyle \frac{d\varepsilon}{dq}\right)}
\left[\frac{d^2 F}{dq^2} - \frac{\displaystyle\frac{d^2 \varepsilon}{dq^2}}{\displaystyle\frac{d\varepsilon}{dq}} \; 
\frac{dF}{dq}\right] \label{Dq2}
\eea
Notice that from eq.(\ref{eqst}) $ \left( \displaystyle -\frac{d\varepsilon}{dq} > 0 \right) $.

Inserting in the integral of eq.(\ref{solabel})
a given expression for the density profile $ F(q) $ and the dimensionless gravitational
potential $ -\varepsilon(q) $ computed from eq.(\ref{eqst}) allows to obtain
the corresponding distribution function $ \Psi $.

\medskip

The coefficient in front of $ {\cal D}(q) $ in eq.(\ref{solabel}) can be evaluated as
\be\label{coefi}
\frac1{G^\frac32 \; r_h^3 \; m^4 \; \sqrt{\rho_0}} = 0.502753 \times 10^{-5} \; \left(\frac{2 \; {\rm keV}}{m}\right)^4 \;
\left(\frac{\rm kpc}{r_h}\right)^\frac52 \; \sqrt{\frac{120 \; M_\odot}{\Sigma_0 \; {\rm pc}^2}} \; .
\ee

For a given profile $ F(q) $, a shift  $ \varepsilon(q) \to \varepsilon(q) + $ constant leaves invariant
the DF $ \Psi(q) $ obtained from eqs.(\ref{solabel})-(\ref{Dq2}).
Taking into account eq.(\ref{defeps}) implies that the DF $ \Psi(q) $  is independent of the value of $ \nu(0) $.

\subsection{Velocity dispersion and Equation of state}\label{vdes}

The average velocity dispersion of the particles depends on $ r $ and follows
from the average momentum as
\be\label{velo}
v^2(r) = \frac1{m^2} \; \frac{\int_0^{\infty} p^4 \; dp \; 
f\left[\displaystyle \frac{p^2}{2m}-\mu(r)\right]}{\int_0^{\infty} p^2 \; dp \; 
f\left[\displaystyle \frac{p^2}{2m}-\mu(r)\right]}
= \frac{2 \, T_0}{m} \; \frac{\int_{\nu(\infty)}^{\nu}  \;  d\nu' \; \left[\nu(q)-\nu' \right]^\frac32 \; \Psi(-\nu')}{
\int_{\nu(\infty)}^{\nu} d\nu' \; \sqrt{\nu(q)-\nu'} \; \Psi(-\nu')} \; .
\ee
From eq.(\ref{den3}) the denominator here is proportional to the density profile and we obtain
\be\label{dispv}
v^2(r) =\frac{2 \, \sqrt2}{\pi^2} \; \frac{m^\frac32 \; T_0^\frac52}{\rho_0 \; F(q)} 
\; \int_{\nu(\infty)}^{\nu} d\nu' \; \left[\nu(q)-\nu' \right]^\frac32 \; \Psi(-\nu') \; .
\ee
We compute this integral in the Appendix A with the result
\be\label{intap}
\int_{\nu(\infty)}^{\nu} d\nu' \; \left(\nu-\nu' \right)^\frac32 \; \Psi(-\nu') = \frac{3 \, \pi^2}{4 \, \sqrt2}
\; \frac{\rho_0}{m^\frac52 \; T_0^\frac32} \; \int_{\nu(\infty)}^\nu d\nu' \; 
\left(\nu-\nu'\right)^2 \; \frac{d^2 F}{d\nu'^2} \; .
\ee
Inserting this expression in eq.(\ref{dispv}) yields the velocity dispersion in terms of
the density profile
\be
v^2(r) = \frac{3 \; T_0}{2 \; m \;  F(q)} \; 
\int_{\nu(\infty)}^\nu d\nu' \; \left[\nu(q)-\nu'\right]^2 \; \frac{d^2 F}{d\nu'^2} \; .
\ee
As above, it is convenient to change the integration variable from $ \nu' $ to $ q' $ with the result
\be \label{vdis}
v^2(r) = 6 \, \pi \; G \; \rho_0 \; r_h^2 \; \frac1{F(q)}
\int_q^{\infty} dq' \; \left[\varepsilon(q)-\varepsilon(q')\right]^2 \; {\cal J}(q')\; .
\ee
where $ {\cal J}(q) $ is given by  eq.(\ref{Dq2}). 
From eq.(\ref{denF}) and eq.(\ref{vdis}) we can immediately compute the pressure in terms of
the density profile $ F(q) $ as 
\bea\label{pres}
&& P(r) = \frac13 \; v^2(r) \; \rho(r) \\ \cr
&& P(r) = 2 \, \pi \; G \; \Sigma_0^2 \; 
\int_q^{\infty} dq' \; \left[\varepsilon(q)-\varepsilon(q')\right]^2 \; {\cal J}(q')\;
\quad , \quad \Sigma_0 \equiv \rho_0 \; r_h \; ,
\eea
where $ \Sigma_0 $ is the surface density.

\medskip

It must be noticed that the surface density  $ \Sigma_0 $
is found nearly {\bf constant} and independent  of 
luminosity in  different galactic systems (spirals, dwarf irregular and 
spheroidals, elliptics) 
spanning over $14$ magnitudes in luminosity and  over different 
Hubble types. More precisely, all galaxies seem to have the same value 
for $ \Sigma_0 $, namely $ \Sigma_0 \simeq 120 \; M_\odot /{\rm pc}^2 $
up to $ 10\% - 20\% $ \citep{dona,span,kor}.   
It is remarkable that at the same time 
other important structural quantities as $ r_h , \; \rho_0 $, 
the baryon-fraction and the galaxy mass vary orders of magnitude 
from one galaxy to another.

{\vskip 0.2cm} 

Given the profile function $ F(q) $, and so the density $ \rho(r) $ from eq.(\ref{denF}),
we determine the pressure $ P(r) $ from eq.(\ref{pres}). In this way, 
we can find the relation between pressure and density associating
to each value of $ \rho(r) $ the corresponding value $ P(r) $. We thus obtain
the equation of state as a numerical table.

{\vskip 0.2cm} 

In subsections \ref{hdmes} and \ref{polyes} below we compute and analyze the halo DM
equation of state and plot it in figs. \ref{psobrerho} and \ref{poly}.

{\vskip 0.2cm} 

The hydro-static equilibrium equation 
\be\label{ehidr}
\frac{dP}{dr} + \rho(r) \; \frac{d\phi}{dr} = 0 \; ,
\ee
is satisfied here as in the case of the Thomas-Fermi theory \citep{newas,astro,dVSS2,nosoct}.

\subsection{The distribution function behaviour near the halo center for cored profiles}\label{dfcqchi}

The lowest energy $ -\varepsilon(q) $ for a DM particle in the galaxy halo occurs
near the center $ q = 0 $. We derive in this subsection
the behaviour of the distribution function for $ q \to 0 $
from the behaviour of cored profile functions $ F(q) $ near the center using 
the explicit formula eq.(\ref{solabel}).

{\vskip 0.2cm} 

For a cored density profile approaching the center as a power of the coordinate we have
in general
\be\label{fqcero}
F(q) \buildrel_{q \to 0}\over= 1 - c \; q^\beta \quad , \quad \beta > 0 \; ,
\ee
where $ c $ and $ \beta $ are positive constants. 

{\vskip 0.2cm} 

For a Burkert profile \cite{burk}
\be\label{perB}
F(q) = \frac1{(q+1)(q^2+1)}  \; ,
\ee
and we have $ c = \beta = 1 $. 

{\vskip 0.1cm} 

A Einasto or Sersic profile can be parametrized as
$$
F(q) = \exp(-q^\beta)
$$
and we have for galaxies $ c = 1 $ and typically $ \beta \sim 0.4 $.

{\vskip 0.1cm} 

Density profiles obtained from primordial density fluctuations \cite{dVSS1} as well as from the 
Thomas-Fermi approach \cite{newas,astro} are even functions of $ r $ and exhibit 
at small distances the behaviour eq.(\ref{fqcero}) with $ \beta = 2 $.

{\vskip 0.2cm} 

From eq.(\ref{eqst}) we find for the particle energy $ -\varepsilon(q) $ near the center $ q = r/r_h \ll 1 $,
\be\label{epsq0}
-\varepsilon(q) \buildrel_{q \to 0}\over= \frac{q^2}6 \left[1 - 
\frac{6 \; c}{(\beta + 2)(\beta + 3)} \; q^\beta\right] \; .
\ee
Inserting eqs.(\ref{fqcero}) and (\ref{epsq0}) in the dimensionless 
distribution function eq.(\ref{solabel}) yields 
\be\label{Dqchico}
 {\cal D}(q)  \buildrel_{q \to 0}\over= \frac{3 \; \sqrt3}{4 \; \sqrt{\pi}} \; c \; \beta
\; (2-\beta) \; \int_q^{\infty} \frac{q'^{\beta-3} \; dq'}{\sqrt{q'^2-q^2}} =
\frac38 \; \sqrt3 \; c \; \beta \; (2-\beta) \; 
\frac{\Gamma\left(\displaystyle\frac{3-\beta}2\right)}{\Gamma\left(2 - \displaystyle\frac{\beta}2\right)} \; 
\frac1{q^{3-\beta}}
\ee
This dimensionless distribution function behaviour can be expressed 
in terms of the particle energy $ -\varepsilon(q) $ with the result
\be\label{Dechico}
{\cal D}(-\varepsilon) \buildrel_{\varepsilon \to 0}\over= \frac{c \; \beta}{8 \; \sqrt2} \; 
\frac{\Gamma\left(\displaystyle\frac{3-\beta}2\right)}{\Gamma\left(1 - \displaystyle\frac{\beta}2\right)} \;
\frac1{(-\varepsilon)^{\frac{3-\beta}2}} \; .
\ee
We see that for $ \beta $ in the interval $ 0 < \beta < 3 $,
the distribution function diverges for the lowest energy $ \varepsilon \to 0 $
except if $ \beta = 2 $. 

{\vskip 0.1cm} 

In particular, eq.(\ref{Dechico}) implies that the DF associated to the  Burkert density profile eq.(\ref{perB})
tends to plus infinity as $ +1/\varepsilon $ for $ \varepsilon \to 0 \; , \; q \to 0 $.

{\vskip 0.1cm} 

In ref. \cite{wid} it is stated that $ {\cal D}(-\varepsilon) $ behaves in the cored
case (bounded $ F(0) $) as $ 1/\varepsilon $ for $ \varepsilon \to 0 $.
This indicates that ref. \cite{wid} choosed $ \beta = 1 $ in eq.(\ref{fqcero}).

{\vskip 0.1cm} 

In addition, from eqs.(\ref{Dqchico})-(\ref{Dechico}) we find
that $ {\cal D}(-\varepsilon) $ for $ \varepsilon \to 0 $ becomes negative (i. e. unphysical) 
for $ \beta $ in the interval:
$$
2 \; l + 1  > \beta > 2 \; l \quad , \quad l = 1,2, 3, \ldots
$$ 

{\vskip 0.2cm} 

The case $ \beta = 2 \; l , \quad l = 1,2, 3, \ldots $ is special. 
For  $ \beta = 2 \; l $ we have
$$
\left. \frac1{\Gamma\left(1 - \displaystyle\frac{\beta}2\right)} \right|_{\beta = 2 \; l} = 0
\quad , \quad l = 1,2, 3, \ldots
$$
and the leading contribution to $ {\cal D}(-\varepsilon) $
in eqs.(\ref{Dqchico})-(\ref{Dechico}) identically vanishes. 

{\vskip 0.2cm} 

Analyzing the $ q \to 0 $ behaviour of the integrand of eq.(\ref{solabel}) we find
that $ {\cal D}(0) $ is {\bf finite and positive} when $ \beta = 2 \; l , \quad l = 1,2, 3, \ldots $.
From eq.(\ref{fqcero}) this implies that physically reasonable density profiles behave for $ q \to 0 $ as
\be\label{Fq2}
F(q) \buildrel_{q \to 0}\over= 1 - c \; q^{2 \; l} \quad , \quad l \geq 1 \; .
\ee
However, a leading $ q \to 0 $ behaviour $ q^{2 \; l} $ with $ l > 1 $ 
describe non-generic density profiles where even powers of $ q $ smaller than $ q^{2 \; l} $ are missing.

{\vskip 0.2cm} 

In summary, only remains the case $ l=1 $ and hence $ \beta = 2 $ in eq.(\ref{fqcero})
describes the generic density profile which provides {\bf finite and positive} distribution
functions near the halo center $ q = r/r_h \ll 1 $:
\be\label{Fqufa}
F(q) \buildrel_{q \to 0}\over= 1 - c \; q^2 
\ee
From now on we shall only consider this cored density profile behaviour $ (\beta = 2) $.

{\vskip 0.2cm} 

We have seen how the analysis of the distribution function near the halo center
allows to select physically meaningful density profiles.

{\vskip 0.2cm} 

The distribution function for spin-$1/2$ fermions must be everywhere smaller than two
in order to satisfy the Pauli principle. Cored density profiles with the behaviour eq.(\ref{Fqufa})
satisfy the Pauli bound as we discuss below in sec. \ref{cotpau}. 

\subsection{The distribution function behaviour near the halo center for cusped profiles}

The generic behaviour of a cusped density profile  near the center $ q = 0 $
is a power-like singularity
\be\label{Fqcusp}
F(q) \buildrel_{q \to 0}\over= \frac{\cal C}{q^\lambda} \quad , \quad \lambda < 3 
\ee
where $ \cal C $ and $ \lambda $ are positive constants. $ \lambda $ must be smaller than 3
to have finite mass.

{\vskip 0.1cm} 

For the NFW profile we have $ \lambda = 1 $ and for the Jaffe profile we have $ \lambda = 2 $ \cite{bin}.

{\vskip 0.2cm} 

From eq.(\ref{eqst}) we find for the particle energy $ -\varepsilon(q) $ near the center $ q = r/r_h \ll 1 $
\be\label{epcu}
-\varepsilon(q) \buildrel_{q \to 0}\over= \frac{\cal C}{(3-\lambda)(2-\lambda)} \; q^{2-\lambda} 
\quad , \quad 0 < \lambda < 3 , \; \lambda \neq 2 \; .
\ee
For $ \lambda = 2 $ we obtain
\be\label{epcu2}
-\varepsilon(q) \buildrel_{q \to 0}\over= {\cal C} \; \ln q + c_1   \to -\infty
\ee
where $ c_1 $ is a constant. We see that
$$
\lim_{q \to 0}\varepsilon(q) = \left\{\begin{array}{l} \displaystyle 0 \quad {\rm for} \quad 0 < \lambda < 2
\; , \\ \\
+\infty \quad {\rm for} \quad 2 \leq \lambda < 3 \; .
\end{array} \right.
$$

Inserting eqs.(\ref{Fqcusp}) and (\ref{epcu}) in the dimensionless 
distribution function eq.(\ref{solabel}) yields near the center $ r =0 $,
\bea\label{Dcuspq}
{\cal D}(q)  &\buildrel_{q \to 0}\over=& \lambda \; (3-\lambda)^\frac32 \; \sqrt{\frac{2-\lambda}{8 \; \pi \; {\cal C}}}
 \; \int_q^{\infty} \frac{dq'}{q'^3 \; \sqrt{q'^{2-\lambda} - q^{2-\lambda}}} =
\frac{\lambda \; (3-\lambda)^\frac32}{\sqrt{8 \; {\cal C} (2-\lambda)}} \; 
\frac{\Gamma\left(\displaystyle\frac12 + \frac2{2-\lambda} \right)}{\Gamma\left(\displaystyle 1 + \frac2{2-\lambda} \right)} \;
\frac1{q^{3 - \lambda/2}} \quad {\rm for} \quad 0 < \lambda < 2 \; , \cr \cr \cr
{\cal D}(q)  &\buildrel_{q \to 0}\over=& \lambda \; (3-\lambda)^\frac32 \; \sqrt{\frac{\lambda-2}{8 \; \pi \; {\cal C}}}
 \; \int_q^{\infty} \frac{dq'}{q'^3 \; \sqrt{q^{2-\lambda} - q'^{2-\lambda}}} =
\frac{\lambda \; (3-\lambda)^\frac32}{\sqrt{8 \; {\cal C} (\lambda-2)}} \; 
\frac{\Gamma\left(\displaystyle \frac2{\lambda-2} \right)}{\Gamma\left(\displaystyle \frac12 + \frac2{\lambda-2} \right)} \;
\frac1{q^{3 - \lambda/2}}\quad {\rm for} \quad 2 < \lambda < 3 \; . \nonumber
\eea
which in terms of the particle energy $ -\varepsilon(q) $ becomes
\bea\label{Dcuspe}
&& {\cal D}(-\varepsilon) \buildrel_{\varepsilon \to 0}\over= \frac{\lambda}{\sqrt8} \; 
\left[\frac{{\cal C}^2}{(3-\lambda)^\lambda}\right]^\frac1{2-\lambda} \; \frac1{(2-\lambda)^{1 + \frac2{2-\lambda}}}
\; \frac{\Gamma\left(\displaystyle \frac12 + \frac2{2-\lambda} \right)}{\Gamma\left(
\displaystyle 1 + \frac2{2-\lambda} \right)} \;
\left(-\varepsilon\right)^{-\frac12 - \frac2{2-\lambda}} \to +\infty
\quad {\rm for} \quad 0 < \lambda < 2 \; , \cr \cr\cr
&& {\cal D}(-\varepsilon) \buildrel_{\varepsilon \to \infty}\over= \frac{\lambda}{\sqrt8} \; 
\left[\frac{(3-\lambda)^\lambda}{{\cal C}^2}\right]^\frac1{\lambda-2} \; (\lambda-2)^{\frac2{\lambda-2}-1} \; 
\frac{\Gamma\left(\displaystyle \frac2{\lambda-2} \right)}{\Gamma\left(\displaystyle \frac12 + \frac2{\lambda-2} \right)} \;
\varepsilon^{\frac{3-\lambda/2}{\lambda-2}} \to +\infty
\quad {\rm for} \quad 2 < \lambda < 3 \; .
\eea
We see that for $  0 < \lambda  < 3 $, the distribution function $ {\cal D}(-\varepsilon) $ {\bf diverges} for 
the lowest energy: for $ \varepsilon \to 0 $ in the case $ 0 < \lambda  < 2 $
and for $ \varepsilon \to \infty $ in the case $ 2 < \lambda < 3 $.

{\vskip 0.1cm} 

Near the center $ q = r/r_h \ll 1 $, we give in  eq.(\ref{Dcuspe}) the power behavior in the energy $ \varepsilon $
and the prefactor for the distribution function $ {\cal D}(-\varepsilon) $ for $ \varepsilon \to 0 $.
We succeded to obtain these results thanks to the formalism eqs.(\ref{solabel})-(\ref{Dq2}).

{\vskip 0.2cm} 

The NFW \cite{nfw} and Hernquist \cite{hern} profiles correspond to $ \lambda = 1 $ and yield
\be\label{NFWH}
 {\cal D}(-\varepsilon) \buildrel_{\varepsilon \to 0}\over= \frac3{32} \; \sqrt{\frac{\pi}2} \; {\cal C}^2 \; \left(-\varepsilon\right)^{-\frac52}
\to +\infty
\ee
The factor $ \left(-\varepsilon\right)^{-\frac52} $ with the exponent $ -5/2 $ 
was found in ref.\cite{wid}.

{\vskip 0.2cm} 

For $ \lambda = 2 $ we obtain from eqs.(\ref{solabel}),  (\ref{Fqcusp}) and (\ref{epcu2})
\be\label{Dcuspq2}
 {\cal D}(q)  \buildrel_{q \to 0}\over=\frac1{\sqrt{2 \; \pi \; {\cal C}}} \; 
\int_q^{\infty} \frac{dq'}{q'^3 \; \sqrt{\ln\left(\displaystyle\frac{q'}{q}\right)}} =
\frac1{\sqrt{2\; {\cal C}} \; q^2} \quad , \quad   \lambda = 2 \; .
\ee
Therefore, we have in terms of the particle energy $ -\varepsilon(q) $
\be\label{Dcue2}
{\cal D}(-\varepsilon) \buildrel_{\varepsilon \to \infty}\over= \frac1{{\cal C}_2} \; 
\exp\left(\displaystyle \frac2{\cal C} \; \varepsilon \right)
\to +\infty \quad , \quad   \lambda = 2 \; .
\ee
where $ {\cal C}_2 $ is a constant. We see that cusped profiles with $ 1/q^2 $ behavior near the center $ q = r/r_h \ll 1 $
produce a Boltzmann distribution function for large energies $ \varepsilon(q) \to +\infty $.
Thermal behavior only appears near the center in the case $ \lambda = 2 $. For $ \lambda \neq 2 $, eqs.(\ref{Dcuspe}) clearly
show a non-Boltzmann behavior for the lowest possible energies.

{\vskip 0.2cm} 

Cusped density profiles yield distribution functions that {\bf diverge} near the halo center
as noticed in ref. \cite{wid,loma,posti}
for specific cusped models (Hernquist, Jaffe, NFW).

{\vskip 0.1cm} 

Clearly, such distribution functions cannot describe spin-$1/2$ fermions since
they violate the Pauli principle near the halo center.

\section{The DM Distribution Function in Galaxies from Empirical halo cored density profiles}

We explicitly compute the phase--space distribution function and the equation
of state for the family of $ \alpha $-density profiles
\be\label{Falfa}
 F(q) = \frac1{\left(1+q^2\right)^\alpha} \quad , \quad 1 \leq \alpha \leq 2.5 \quad , \quad 
q = \frac{r}{r_h} \; .
\ee
This cored density profile is a generalization of the pseudo-thermal profile and it is perfectly
appropriate to fit galaxy observations. $ r_h $ is defined as the point where the density profile
takes the value
\be\label{rhalfa}
F(r  = r_h) = 2^{-\alpha} \; .
\ee
For the profiles eq.(\ref{Falfa}), the value of $\alpha $  must be 
chosen before fitting the data. Notice that the halo radius 
$ r_h $ defined by  eq.(\ref{rhalfa}) turns to be $\alpha $-dependent.

{\vskip 0.2cm} 

As shown in subsection \ref{dfcqchi}, density profiles with the short distance behaviour
eq.(\ref{Fqufa}) produce meaningful distribution functions at short distances.
The $\alpha $-density profiles eq.(\ref{Falfa}) fulfil eq.(\ref{Fqufa}) at small distances.
This is not the case of Burkert, Einasto and Sersic profiles
which exhibit divergent distribution functions at short distances 
[see eqs.(\ref{Dqchico})-(\ref{Dechico}) for $ \beta \neq 2 $].
For all these reasons we will concentrate from now on 
the family of density profiles given by eq.(\ref{Falfa}).

{\vskip 0.1cm} 

An $\alpha $-density profile eq.(\ref{Falfa}) with $ \alpha = 1.5913 $
is an excellent approximation to the Thomas-Fermi density profile in the internal region
$ r \lesssim 3 \; r_h $ \cite{nosoct}.

{\vskip 0.1cm} 

Observations for all $ r $ till the virial radius are best fitted for the values
$$ 
\alpha = 1.2 - 1.5 \; .
$$ 

In particular, for $ q \gg 1 $ and $ \alpha = 1.5 $, eq.(\ref{Falfa}) reproduces
the Burkert density profile.

{\vskip 0.1cm} 

Spiral and elliptic galaxy observations favour the value $ \alpha = 1.2 $ \cite{paol}. Although
observations disfavour $ \alpha $ values outside the  1.2 - 1.5 range, it is useful 
and illustrative for the purposes of understanding to analyze 
the whole range of values $ 1 < \alpha \leq 2.5 $.

{\vskip 0.1cm} 

The value $ \alpha = 2.5 $ is not appropriate to describe realistic DM halos but corresponds to the
Plummer profile describing the density of stars in globular clusters \cite{plu,bin}.

{\vskip 0.1cm} 

For $ \alpha = 1.5 $, eq.(\ref{Falfa})  becomes the modified Hubble model \cite{bin}.

\medskip

\begin{figure}
\begin{turn}{-90}
\psfrag{"RLtalfa2.dat"}{$ \alpha = 1.1 $}
\psfrag{"Ltalfa3.dat"}{$ \alpha = 1.2 $}
\psfrag{"Ltalfa4.dat"}{$ \alpha = 1.3 $}
\psfrag{"Ltalfa5.dat"}{$ \alpha = 1.4 $}
\psfrag{"Ltalfa32.dat"}{$ \alpha = 1.5 $}
\psfrag{"Ltalfa10.dat"}{$ \alpha = 1.6 $}
\psfrag{"Ltalfa12.dat"}{$ \alpha = 1.7 $}
\psfrag{"Ltalfa14.dat"}{$ \alpha = 2 $}
\psfrag{"Ltalfa25.dat"}{$ \alpha = 2.5 $}
\psfrag{"trucho.dato"}{}
\psfrag{-\varepsilon(q)}{$-\varepsilon(q)$}
\psfrag{q = \log r/r_h}{$ \log q = \log r/r_h $}
\includegraphics[height=12.cm,width=10.cm]{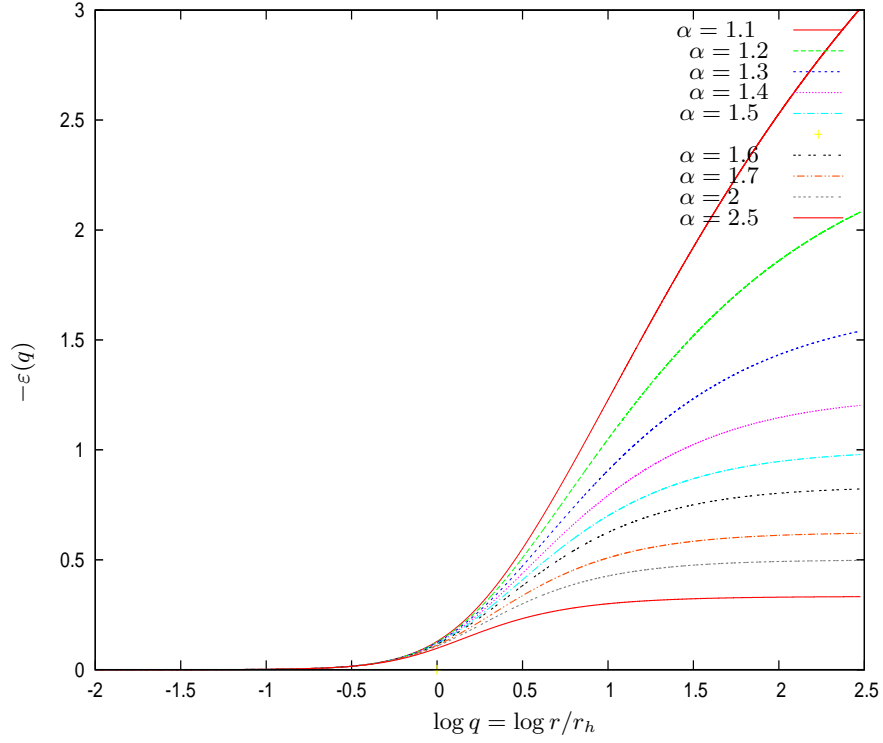}
\end{turn}
\caption{The dimensionless energy $ -\varepsilon(q) $ given by eq.(\ref{tq}) 
vs. the ordinary logarithm of the coordinate $ q = r/r_h $ for relevant values of the exponent $ \alpha $
in the density profile eq.(\ref{Falfa}). For increasing $ q , \; -\varepsilon(q) $ increases slowly.
Notice the {\bf shallow} potential felt by the particles inside the halo for $ r<r_h , \; \log_{10} q < 0 $.}
\label{tdeq}
\end{figure}

For the density profile eq.(\ref{Falfa}), the normalized gravitational potential
$ \varepsilon(q) $ eq.(\ref{eqst}) is given by
\bea\label{tq}
&& \varepsilon(q) = \frac{w_\alpha(q)}{q} + 
 \frac1{2 \, (\alpha-1)} \left[\frac1{\left(1+q^2\right)^{\alpha-1}} - 1\right] 
\quad , \quad \varepsilon(\infty) = -\frac1{2 \, (\alpha-1)} \; ,
\\ \cr \cr
&& w_\alpha(q) \equiv \int_0^q \frac{s^2 \; ds}{\left(1+s^2\right)^\alpha} 
=\frac{q^3}3 \;  {}_2F_1\left(\alpha,\frac32;\frac52; -q^2\right)
\; , \label{walfa}
\eea
where $ {}_2F_1\left(a,b;c;z\right) $ stands for the hypergeometric function \cite{gra}.
At fixed $ q , \; w_\alpha(q) $ monotonically decreases when $ \alpha $ grows.

{\vskip 0.2cm} 

\begin{figure}
\begin{turn}{-90}
\psfrag{"epsalfa.dat"}{$ -\varepsilon(q=1) $ vs. $ \alpha $}
\psfrag{-\varepsilon(1)}{$-\varepsilon(1)$}
\psfrag{alpha}{$\alpha$}
\includegraphics[height=12.cm,width=10.cm]{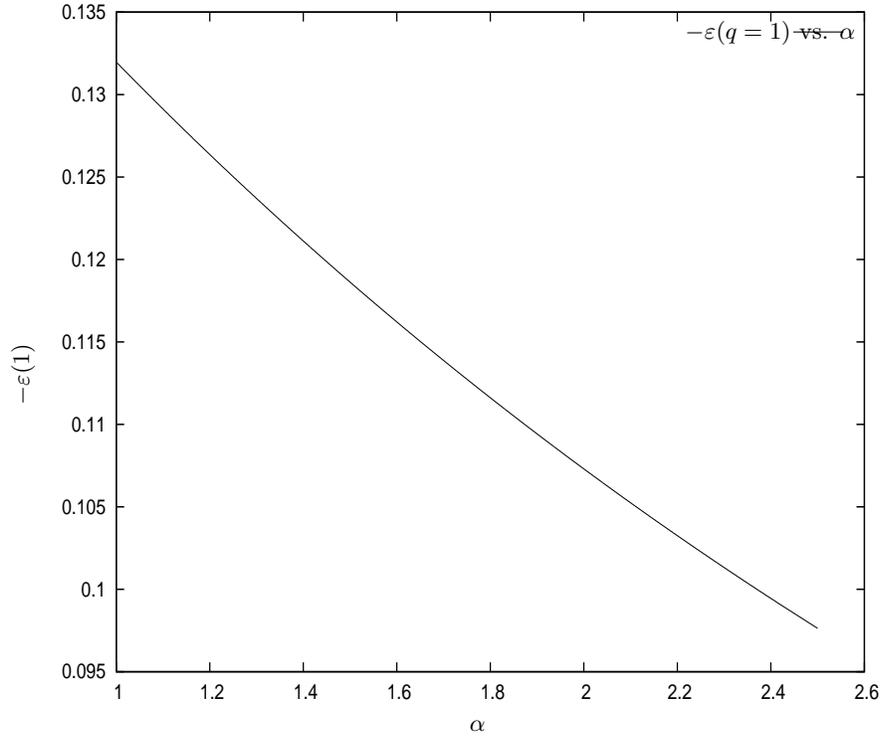}
\end{turn}
\caption{The dimensionless energy at the halo radius $ r = r_h, \; -\varepsilon(1) $ vs. the exponent $ \alpha $
in the density profile eq.(\ref{Falfa}). $ -\varepsilon(q=1) $ turns to be near 0.1 in the whole 
range $ 1 < \alpha < 2.5 $. That is, the potential energy of a particle is quite small inside the halo,
$ q < 1 $.}
\label{epsalfa}
\end{figure}

In the particular cases $ \alpha = 1, \frac32 , \; 2 , \; \frac52 $ the function $ w_\alpha(q) $ reduces
to elementary functions \cite{pru}:
$$
w_1(q) = q - \arctan q \quad , \quad \varepsilon(q) =1 - \frac{\arctan q}{q}-\frac12 \; \ln\left(1+q^2\right) 
\quad , \quad  \alpha = 1\; ,
$$
\be\label{w15}
w_{\frac32}(q) = {\rm Arg \, Sinh} (q) - \frac{q}{\sqrt{1+q^2}} \quad , \quad 
\varepsilon(q) = \frac{{\rm Arg \, Sinh} (q)}{q} - 1 \quad , \quad \alpha = \frac32 \; ,
\ee
$$
w_2(q) =\frac{q}2 \; \left(\frac{\arctan q}{q} - \frac1{1+q^2}\right)
\quad , \quad 
\varepsilon(q) = -\frac12 \; \left(1 - \frac{\arctan q}{q}\right) \quad , \quad \alpha = 2 \; ,
$$
$$
w_{\frac52}(q) = \frac{q^3}{3 \, \left(1 + q^2 \right)^{\frac32}} \quad , \quad 
\varepsilon(q) = -\frac13 \; \left(1 -\frac1{\sqrt{1+q^2}}\right) \quad , \quad \alpha =\frac52 \; .
$$
We plot in fig. \ref{tdeq} the dimensionless energy $ -\varepsilon(q) $ as a function of $ \log_{10} q $
for relevant values of the exponent $ \alpha $ in the density profile eq.(\ref{Falfa}). 
We see that $ -\varepsilon(q) $ grows monotonically with $ q $. Small energy values $ -\varepsilon(q) \lesssim 0.1 $
are confined inside the halo radius $ r \lesssim r_h $, i. e. $ q \lesssim 1 $. 

\medskip

We plot in fig. \ref{epsalfa} the energy at the halo radius $ r = r_h, \; -\varepsilon(q=1) $ as a function of 
$ \alpha $ in the interval $ 1 < \alpha < 2.5 $. We see that $ -\varepsilon(1) \simeq 0.1 $
in this range. In particular, $ -\varepsilon(1) = 0.118626 $ for $ \alpha = \frac32 $.

{\vskip 0.1cm} 

As a consequence, only particles with $ -\varepsilon(q) \lesssim 0.1 $ stay inside the halo $ r < r_h $.
$ -\varepsilon(q) $ grows slowly when the distance to the origin $ q $ grows.
For example, $ -\varepsilon(10) = 0.700178 $ for $ \alpha = \frac32 $
as shown in fig. \ref{tdeq}.

\subsection{The Galaxy Halo Mass and Scaling Relations}\label{ghmsr}

The dark matter mass inside a radius $ R $ follows by integrating the mass density eqs.(\ref{denF}) and (\ref{Falfa})
\be\label{masar}
M(r) =  4 \, \pi \int_0^{r} r'^2 \; dr' \; \rho(r') = 4 \; \pi \; \rho_0 \; r_h^3 \;  w_\alpha(q) \; .
\ee
For $ r \to \infty , \; M(r) $ has a bounded limit for $ \alpha > \frac32 $.
The $ q = \infty $ limit of $ w_\alpha(q) $ follows from eqs.(\ref{walfa}) with the result\cite{gra}
$$
w_\alpha(\infty) = \frac{\sqrt{\pi}}4 \; 
\frac{\Gamma\left(\alpha - \frac32\right)}{\Gamma\left(\alpha\right)} 
\quad , \quad \alpha > \frac32 \; .
$$
We find in the particular cases $ \alpha = 2 , \; \frac52 $
$$
w_2(\infty) = \frac{\pi}4  \quad , \quad w_\frac52(\infty) = \frac13 \quad .
$$

The halo mass $ M_h $ follows by setting here $ r = r_h $ (i. e. $ q = 1 $):
\be\label{mhd}
M_h \equiv M(r_h) = 4 \, \pi \int_0^{r_h} r^2 \; dr \; \rho(r) =  d_\alpha \; \rho_0 \; r_h^3 
\quad , \quad d_\alpha \equiv 4 \; \pi \;  w_\alpha(1) \; ,
\ee
where $ w_\alpha(q=1) $ is given by eq.(\ref{walfa}). It follows from eq.(\ref{walfa})
that $ w_\alpha(1) $ is a decreasing function of $ \alpha $.

{\vskip 0.2cm} 

This scaling law eq.(\ref{mhd}) is a direct consequence of the assumed density profile
eq.(\ref{denF}). The value of $ d_\alpha $ depends on the choice of the function $ F(q) $.

{\vskip 0.2cm} 

In the cases $ \alpha = 1, \frac32 , \; 2 , \; \frac52 , \;  d_\alpha $ takes the values
$$
d_1 = \pi (4 - \pi) = 2.696766 \quad , \quad 
d_{\frac32} = 4 \; \pi \; \left[ \ln\left(1 + \sqrt2 \right) - \frac1{\sqrt2} \right] = 2.18990 
$$
$$
d_2 = \frac{\pi}2 \; (\pi - 2 ) = 1.793210 \quad , \quad d_{\frac52} = \frac{\pi}3 \; \sqrt2 =1.480961 \; .
$$

Both the amplitude and exponent in eq.(\ref{mhd}) are well verified by 
observations on large enough galaxies in the dilute
regime: $ M_h > 10^6 \; M_\odot $.

{\vskip 0.2cm} 

It must be stressed that the Thomas-Fermi (TF) theoretical approach to galaxy structure
yields a similar formula to eq.(\ref{mhd}) for the halo mass $ M_h $ of galaxies in the dilute
regime ($ M_h > 10^6 \; M_\odot $) \cite{nosoct}
$$
M_h = d_{TF} \; \rho_0 \; r_h^3 \quad , \quad d_{TF} = 1.75572 \; .
$$
On the other hand, the empiric Burkert profile \cite{burk} yields a similar expression for $ M_h $
with $ d_{Burkert} = 1.59796 $ \cite{newa}.

{\vskip 0.2cm} 

The slight difference between the coefficients $ d $ for the different types of profiles
can be retraced from the fact that the shapes of the Thomas-Fermi profile, 
Burkert profile, and the $ \alpha $-profiles eq.(\ref{Falfa}) are different.
Consequently, for a given halo mass $ M_h $,
the halo radius eq. (\ref{onequarter}) for these different profiles are
different to each other, and the coefficients $ d $ are slightly different too.

{\vskip 0.1cm} 

It is remarkable, however, that all these coefficients $ d $ are close to each other
indicating the {\bf robustness} of the scaling relation 
\be\label{mscal}
M_h = d \; \rho_0 \; r_h^3 = d \; \Sigma_0 \;  r_h^2 \; ,
\ee
where $ M_h $ is defined by eq.(\ref{mhd}).

\subsection{The Dark Matter Potential Energy}

The dark matter potential energy inside a radius $ R $ is given by
$$
U(R) = - 4 \; \pi \; G \int_0^R r \; dr \; \rho(r) \; M(r) \quad .
$$
Using eq.(\ref{denF}) and eq.(\ref{masar}) yields
\be\label{uQ}
U(R) =  - (4 \; \pi)^2 \; G \; \Sigma_0^2 \; r_h^3 \; u(Q) \quad , \quad 
u(Q) \equiv \int_0^Q q \; dq \; F(q) \; w_\alpha(q) \quad , \quad Q \equiv R/r_h \; .
\ee
At fixed $ Q , \; u(Q) $ monotonically decreases for increasing $ \alpha $.

{\vskip 0.1cm} 

For $ R \to \infty , \; U(R) $ has a bounded limit for $ \alpha > 1.25 $.
The $ Q = \infty $ limit of $ u(Q) $ follows from eqs.(\ref{walfa}) and (\ref{uQ}) with the result \cite{gra}
\be\label{uinf}
u_\alpha(\infty) = \frac{\sqrt{\pi}}{8\, (\alpha-1)} \; 
\frac{\Gamma\left(2 \,\alpha - \frac52\right)}{\Gamma\left(2 \,\alpha - 1\right)} 
\quad , \quad \alpha > \frac54 \; .
\ee
In particular, for $ \alpha = \frac32 , \; 2 , \; \frac52 $ we have
$$
u_\frac32(\infty) = \frac{\pi}4 \quad , \quad u_2(\infty) = \frac{\pi}{32} \quad , \quad
u_\frac52(\infty) =\frac{\pi}{96}\; .
$$

\subsection{The resulting distribution function from the Eddington equation turns to be locally thermal.}\label{delt}

The distribution function follows by inserting the density profile eq.(\ref{Falfa}) 
and the potential energy $ \varepsilon(q) $ eq.(\ref{tq}) in the integral representation eq.(\ref{solabel})
with the result
\bea
&& \Psi(q) = \frac1{G^\frac32 \; r_h^3 \; m^4 \; \sqrt{\rho_0}} \; \; {\cal D}(q) \quad ,
\\ \cr 
&& {\cal D}(q) \equiv \frac{\alpha}{\sqrt{8 \, \pi}} \; 
\int_q^{\infty} \frac{q'^2 \; dq'}{\sqrt{\varepsilon(q)-\varepsilon(q')}} \;\frac1{w_\alpha^2(q') \; 
\left(1+q'^2\right)^{\alpha+2}}
\; \left\{\left[(2 \, \alpha -1) \; q'^2 -3 \right] \; w_\alpha(q') + 
\frac{q'^3}{\left(1+q'^2\right)^{\alpha-1}}\right\} \; .\label{Dq}
\eea

\medskip

\begin{figure}
\begin{turn}{-90}
\psfrag{"yLDalfa1.dat"}{$ \alpha = 1.4 $}
\psfrag{"yLDalfa2.dat"}{$ \alpha = 1.509 $}
\psfrag{"yLDalfa3.dat"}{$ \alpha = 1.6 $}
\psfrag{-\varepsilon(q)}{$-\varepsilon(q)$}
\psfrag{{\cal D}(-\varepsilon(q))}{${\ln \cal D}(-\varepsilon(q))$}
\includegraphics[height=12.cm,width=10.cm]{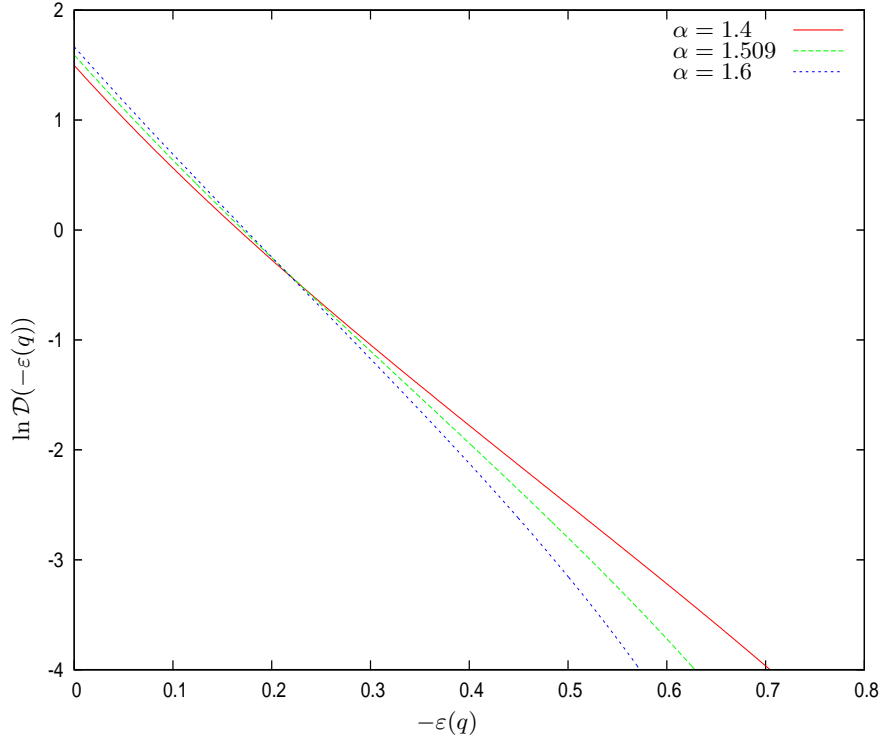}
\end{turn}
\caption{The natural logarithm of the normalized distribution function $ {\cal D}(-\varepsilon) $ eq.(\ref{Dq})
vs. the energy $ -\varepsilon $ eq.(\ref{tq}) for the values of the 
exponent $ \alpha $ in the density profiles eq.(\ref{Falfa})
$ \alpha = 1.4, \; 1.509 $ and $ 1.6 $ appropriate for DM halos.
Notice the {\bf linear} behaviour of $ \ln {\cal D}(-\varepsilon) $ indicating an approximately {\bf Boltzman}
distribution function $ {\cal D}(-\varepsilon) $ (and approximate thermalization) for $ 0 \leq -\varepsilon \lesssim 0.7 $ and 
$ 0 < q \lesssim 7 $, i. e. $ 0 < r \lesssim 7 \; r_h $.}
\label{lnD}
\end{figure}

From eqs.(\ref{vdis}) and (\ref{pres}) we obtain analogous expressions for the
velocity dispersion and the pressure
\bea\label{vepre}
&& v^2(r) = 3 \; G \; \Sigma_0 \; r_h \; \frac{\Pi(q)}{F(q)} \quad , \quad 
P(r) = G \; \Sigma_0^2 \; \Pi(q) \; ,\\ \cr 
&& \Pi(q) \equiv 4 \, \pi \; \alpha \; \int_q^{\infty} q'^2 \; dq' \; 
\frac{\left[\varepsilon(q)-\varepsilon(q')\right]^2}{w_\alpha^2(q') \; \left(1+q'^2\right)^{\alpha+2}}
\; \left\{\left[(2 \, \alpha -1) \; q'^2 -3 \right] \; w_\alpha(q') + 
\frac{q'^3}{\left(1+q'^2\right)^{\alpha-1}}\right\} \; .\label{defPI}
\eea
The hydrostatic equilibrium equation eq.(\ref{ehidr}) in dimensionless 
variables takes the form
\be\label{hidsd}
\frac{d\Pi}{dq} = 4 \; \pi \; F(q) \; \frac{d\varepsilon}{dq} = 
-4 \; \pi \; \frac{F(q)}{q^2} \; \int_0^q q'^2 \;  F(q') \; dq'\; .
\ee

In the spherically symmetric case, the pressure and the velocity dispersion 
can be obtained integrating the hydrostatic equilibrium equation eq.(\ref{ehidr})
instead of using eqs.(\ref{velo})-(\ref{pres}). Also, in the spherically symmetric case,
the Jean's equation from which the pressure and the velocity dispersion follow, reduce to 
the hydrostatic equilibrium equation \cite{bin}.

{\vskip 0.2cm} 

In the particular cases $ \alpha = 1, \frac32 , \; 2 , \; \frac52 $ the 
hydrostatic equilibrium equation eq.(\ref{hidsd}) can be integrated in close form with the 
following results:
$$
\Pi(q) = 4 \; \pi \left[\frac{\pi^2}8 - \frac{\arctan q}{q} - \frac12 \; \left(\arctan q\right)^2 \right]
\quad , \quad  \alpha = 1\; ,
$$
\be\label{pi15}
\Pi(q) = 4 \; \pi \left[\frac1{2(1+q^2)} + \frac{1+ 2 \; q^2}{q \; \sqrt{1+q^2}} \; {\rm Arg \, Sinh}(q)
- \ln(1+q^2) - \ln 4 \right]\quad , \quad  \alpha = \frac32 \; ,
\ee
$$
\Pi(q) = \frac{\pi}2 \left[\frac{3 \; q^2 + 4}{(1+q^2)^2} + \frac{2(3 \; q^2 + 2)}{q \; (q^2+1)} \; \arctan q
+ 3 \; \left(\arctan q\right)^2 - \frac34 \; \pi^2 \right]
\quad , \quad  \alpha = 2 \; ,
$$
$$
\Pi(q) = \frac{2 \; \pi}9 \frac1{(1+q^2)^3} \quad , \quad  \alpha = \frac52 \; .
$$
For the Plummer profile of globular star clusters ($ \alpha = \frac52 $) it follows that
a polytropic equation of state is fulfilled exactly
\be\label{poly25}
\Pi(q) = \frac{2 \; \pi}9 \; F^{\frac65}(q) \quad , {\rm i. e. } , \quad P(r) = \frac{2 \; \pi}9 \;
G \; \Sigma_0^2 \; \left[ \frac{\rho(r)}{\rho_0}\right]^{\frac65} \; .
\ee

\begin{figure}
\begin{turn}{-90}
\psfrag{"yLDalfa4.dat"}{$ \alpha = 2.5 $ (Plummer profile)}
\psfrag{-\varepsilon(q)}{$-\varepsilon(q)$}
\psfrag{{\cal D}(-\varepsilon(q))}{$\ln{\cal D}(-\varepsilon(q))$}
\includegraphics[height=12.cm,width=10.cm]{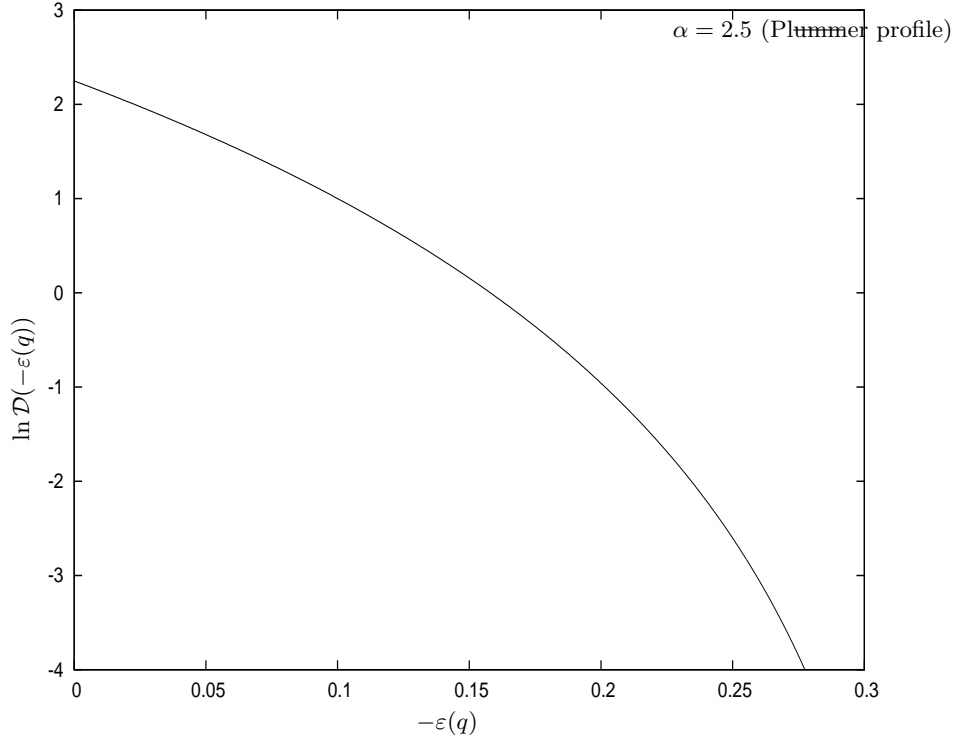}
\end{turn}
\caption{The natural logarithm of the normalized distribution function $ {\cal D}(-\varepsilon) $ eq.(\ref{Dq})
vs. the dimensionless energy $ -\varepsilon $ eq.(\ref{tq}) for the value $ \alpha = 2.5 $ which
corresponds to the Plummer profile for stellar globular clusters. The distribution function $ {\cal D}(-\varepsilon) $
is seen to be approximately of Boltzmann type only in the small range $  0 \leq -\varepsilon \lesssim 0.07 $ 
corresponding to $ 0 \leq q < 0.78 $.}
\label{plum}
\end{figure}

We plot in fig. \ref{lnD} the natural logarithm of the normalized distribution function 
$ {\cal D}(-\varepsilon) $ eq.(\ref{Dq}) vs. $ -\varepsilon $ for the values $ \alpha = 1.4, \; 1.509 $ and $ 1.6 $,
 appropriate for DM halos. Notice the {\bf linear} behaviour of $ \ln {\cal D}(-\varepsilon) $ 
with $ -\varepsilon $ which thus indicates an approximately Boltzmann distribution function $ {\cal D}(-\varepsilon) $.
We find that:

\begin{itemize}
\item{$ {\cal D}(q) $ and thus $ \Psi(q) $ are {\bf positive} for all values of $ q $ in the whole range 
$ 1 \leq \alpha \leq 2.5 $. This shows that the density profiles eq.(\ref{Falfa}) are physically meaningful.
Notice that in general there is no guarantee that $ \Psi(q) $ from eq.(\ref{solabel}) 
will be nowhere negative \cite{bin}.}
\item{$ \ln {\cal D}(-\varepsilon) $ is approximately a {\bf linear} function of the 
energy $ -\varepsilon $ for $ \alpha \sim 1.5 $ in the range of energies $ 0 < -\varepsilon \lesssim 0.6 $ 
which corresponds to $ 0 < q \lesssim 7 $, that is $ 0 < r \lesssim 7 \; r_h $ as can be seen from fig. \ref{lnD}. 
Therefore, the distribution function $ {\cal D}(-\varepsilon) $ is 
approximately a {\bf thermal Boltzman} distribution function in this interval.}
\item{The distribution function $ {\cal D}(-\varepsilon) $ monotonically decreases for growing energy 
$ -\varepsilon $ as it should be, as can be seen from fig. \ref{lnD}.
At fixed $ \varepsilon ,  \; {\cal D}(-\varepsilon) $ increases for decreasing $ \alpha $.
That is, for decreasing $ \alpha $ the density profile is shallower,
particles are less gravitationally bounded and particle states with higher energy are populated.}
\item{The maximum value of $ {\cal D}(-\varepsilon) $ is at $ \varepsilon = 0 $. 
For density profiles appropriate for DM halos ($ \alpha \sim 1.5 $) we have $ {\cal D}(0) < 6 $
(see fig. \ref{lnD}). For diluted galaxies with $ M_h > 10^6 \; M_\odot , \; r_h > 100 $ pc,
we find from eqs.(\ref{solabel}) and (\ref{coefi}) that the distribution function $ \Psi $ is smaller than unity
indicating a classical regime. Besides, the Pauli bound for fermionic DM, $ \Psi \leq 2 $, is fulfilled.}
\item{For comparison we depict in fig. \ref{plum} the distribution function for the Plummer 
(stellar globular cluster) profile ($ \alpha = 2.5 $).
We see that it is approximately thermal only for a small range of energies $ 0 \leq -\varepsilon \lesssim 0.07 $, 
namely, for $ 0 \leq q \lesssim 0.78 $. Therefore, stars in globular clusters are approximately thermal
in a narrower region both in energy and coordinates than the DM in galaxy halos.}
\end{itemize}

{\vskip 0.2cm} 

\begin{figure}
\begin{turn}{-90}
\psfrag{"rLDalfa1.dat"}{$ \alpha = 1.4 $}
\psfrag{"rLDalfa2.dat"}{$ \alpha = 1.509 $}
\psfrag{"rLDalfa3.dat"}{$ \alpha = 1.6 $}
\psfrag{-\varepsilon(q)}{$-\varepsilon(q)$}
\psfrag{{\cal D}(-\varepsilon(q))}{$\ln{\cal D}(-\varepsilon(q))$}
\includegraphics[height=12.cm,width=10.cm]{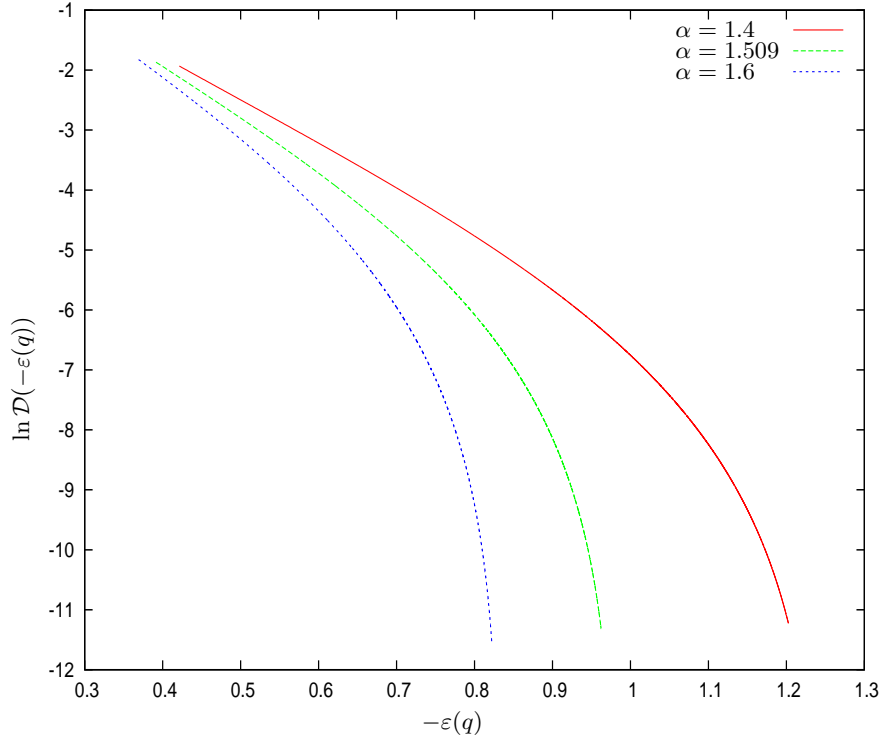}
\end{turn}
\caption{The natural logarithm of the normalized distribution function $ {\cal D}(-\varepsilon) $ eq.(\ref{Dq})
outside the halo radius $ r > 3 \; r_h $ vs. the energy $ -\varepsilon $ eq.(\ref{tq}) 
for the DM halo density profiles eq.(\ref{Falfa}) $ \alpha = 1.4, \; 1.509 $ and $ 1.6 $.
This figure complements fig. \ref{lnD}. $ \ln {\cal D}(-\varepsilon) $ vs. $ -\varepsilon $ 
is approximately a straight line with slope $ 1/t_0 $ for $ 0 \leq -\varepsilon \lesssim 0.7 $ and 
$ 0 < q \lesssim 7 $, i. e. $ 0 < r \lesssim 7 \; r_h $ which indicates a Boltzmann distribution
with temperature $ T_0 $. We find approximate thermalization in this interval.
The slope $ 1/t(q) $ [eq.(\ref{bolt})] of $ \ln {\cal D}(-\varepsilon) $ vs. $ -\varepsilon $ 
increases for increasing $ -\varepsilon $ indicating that the local temperature 
$ T(r) $ decreases with $ r $ for $ -\varepsilon \gtrsim 0.7 $.}
\label{rlnD}
\end{figure}

Therefore, dark matter described by a cored density profile as eq.(\ref{Falfa}) is {\bf approximately
in thermal equilibrium} for $ r \lesssim 7 \; r_h $. It must be recalled that
empiric cored density profiles as eq.(\ref{Falfa}) are good approximations
to real observational data especially for $ \alpha \sim 1.5 $. 

{\vskip 0.1cm} 

Notice that the density profiles apply within the virial radius $ R_{virial} $ whose
typical values are $ 10 \; r_h \lesssim R_{virial} \lesssim 100 \; r_h $
and that the distribution function becomes very small for $ q \gtrsim 10, \; (r \gtrsim 10 \; r_h)  $.

{\vskip 0.2cm} 

In the regions where the distribution function $ {\cal D}(-\varepsilon) $ is approximately Boltzmann-like
we can write it as
\be\label{bolt}
\ln {\cal D}(-\varepsilon) = -\frac{E(q)}{T_0} -A_0 \; ,
\ee
where $ E(q) $ [eq.(\ref{Eq})] is the potential energy of a particle at the point $ q $,
$ T_0 $ the temperature and $ A_0 $ is a normalization constant. We have in dimensionless variables,
\be\label{bolt2}
\ln {\cal D}(-\varepsilon) = \frac{\varepsilon(q)}{t_0} -A_0 \; ,
\ee
where the dimensionless temperature $ t_0 $ is given by
\be\label{tcero}
 t_0 = \frac1{4 \; \pi \; G \; \Sigma_0 \; r_h } \; \frac{T_0}{m} = \frac1{b_0} \; .
\ee
The coefficient $ b_0 $ is defined in eq.(\ref{defb}) and 
\be\label{GSR}
4 \; \pi \; G \; \Sigma_0 \; r_h = 7.21624 \; 10^{-8} \; \frac{r_h}{\rm kpc} \; 
\frac{\Sigma_0 \; {\rm pc}^2}{120 \; M_\odot} \; .
\ee
From a least square fit to the numerical integration of $ {\cal D}(-\varepsilon) $
eq. (\ref{Dq}) for $ \alpha = 1.509 $ in the range displayed in fig. \ref{lnD}, we obtain
$$
a_0 = 1.53 \quad , \quad b_0 = 8.72 \quad ,
$$
and therefore, we find for the dimensionless temperature $ t_0 $,
\be\label{fiteo}
t_0 = 0.115 = {\cal O}(0.1) \; .
\ee
which corresponds to
\be
\frac{T_0}{m} \sim 10^{-8} \; \frac{r_h}{\rm kpc} \; \frac{\Sigma_0 \; {\rm pc}^2}{120 \; M_\odot} \; .
\ee
More precisely,
\be\label{T0K}
T_0 = 1.675 \; t_0 \; \frac{m}{2 \; {\rm keV}} \; \frac{r_h}{\rm kpc} \; 
\frac{\Sigma_0 \; {\rm pc}^2}{120 \; M_\odot} \; {\rm K} = 0.192 \; \frac{m}{2 \; {\rm keV}} \; \frac{r_h}{\rm kpc} \; 
\frac{\Sigma_0 \; {\rm pc}^2}{120 \; M_\odot} \; {\rm K} \; .
\ee
For $ m $ in the keV scale (warm dark matter), $ T_0 $ runs from the mili-Kelvin
for dwarf galaxies till tens of Kelvin for the largest galaxies \cite{nosoct}.

{\vskip 0.2cm}

We plot in fig. \ref{rlnD} the natural logarithm of the distribution function 
$ {\cal D}(-\varepsilon) $ eq.(\ref{Dq}) outside the halo radius $ r > 3 \; r_h $ for  
$ \alpha = 1.4, \; 1.509 $ and $ 1.6 $ vs. the energy $ -\varepsilon $.
We see that the slope $ 1/t(q) $ of $ \ln{\cal D}(-\varepsilon) $ vs. $ -\varepsilon $ 
increases for increasing $ -\varepsilon $. From eqs.(\ref{bolt}) and (\ref{tcero}),
an increasing slope implies a decreasing temperature
$ t(q) $ outside $ r = 3 \; r_h $ while $ t(q) = t_0 $ is constant  inside 
$ r = r_h $. This suggests that $ t(q)  $, and therefore $ T(r) $, can be considered
a {\bf local temperature} that slowly decrease with $ r $ for $ r > 3 \; r_h $.

{\vskip 0.1cm}

As shown in sec. \ref{eqtloc} below, the local temperature decrease with $ r $ for $ r > 3 \; r_h $
is similar when $ T(r) $ is derived from the equation of state (obtained from the velocity dispersion).

\subsection{The Pauli bound on the distribution function}\label{cotpau}

Expressing the halo radius $ r_h $ in terms of the halo mass $ M_h $ in eq.(\ref{coefi}) 
with the help of the scaling relation eq.(\ref{mscal}) yields for the distribution function
\be\label{PsiD}
 \Psi(q) = 0.00531939 \; \left(\frac{10^6 \; M_\odot}{M_h}\right)^\frac54 \; 
\left(\frac{\Sigma_0 \; {\rm pc}^2}{120 \; M_\odot}\right)^\frac34 \; 
\; \left(\frac{2 \; {\rm keV}}{m} \right)^4 \; {\cal D}(q) \; .
\ee
The DF derived in the classical regime
of the collisionless self-gravitating DM must obey the quantum constraints
arising from the quantum nature of the DM particles.
The Pauli bound for spin-$1/2$ DM particles requests that $ \Psi(q) \leq 2 $.
The distribution function takes its maximum value at the halo center $ q = 0 $
where $ {\cal D}(0) = 4.914245 $ for $ \alpha = 1.509 $.
Thus, from eq.(\ref{PsiD}), the Pauli bound is everywhere satisfied provided
the halo mass is bounded from below as
$$
M_h \geq 3.11 \; 10^4 \; M_\odot \; \left(\frac{\Sigma_0 \; {\rm pc}^2}{120 \; M_\odot}\right)^{\! \frac35} \; 
\left(\frac{2 \; {\rm keV}}{m}\right)^{\! \frac{16}5} \; .
$$
This lower bound corresponds to quantum fermions in the degenerate limit  \cite{nosoct}
showing the consistency of the distribution function obtained from the Eddington
equation with the quantum treatment of the DM in ref. \cite{astro,dVSS1,dVSS2,nosoct}.

{\vskip 0.2cm} 

Indeed, baryons may play a role in the halo properties near $ r = 0 $ and in shaping $ \rho(r) $ there.
However, the DF  eq.(\ref{PsiD}) and the  lower bound in the degenerate limit  \cite{nosoct}
both correspond to pure DM and it is therefore consistent to compare them.
 
\section{The Halo Dark Matter equation of state}\label{hdmes}

\begin{figure}
\begin{turn}{-90}
\psfrag{"Npsobrerhoqalfa11.dat"}{$ \alpha = 1.4 $}
\psfrag{"Npsobrerhoqalfa12.dat"}{$ \alpha = 1.509 $}
\psfrag{"Npsobrerhoqalfa5.dat"}{$ \alpha = 1.6 $}
\psfrag{q = r/r_h}{$ q = r/r_h $}
\psfrag{\log t(q)}{$\log_{10} t(q)$}
\includegraphics[height=12.cm,width=10.cm]{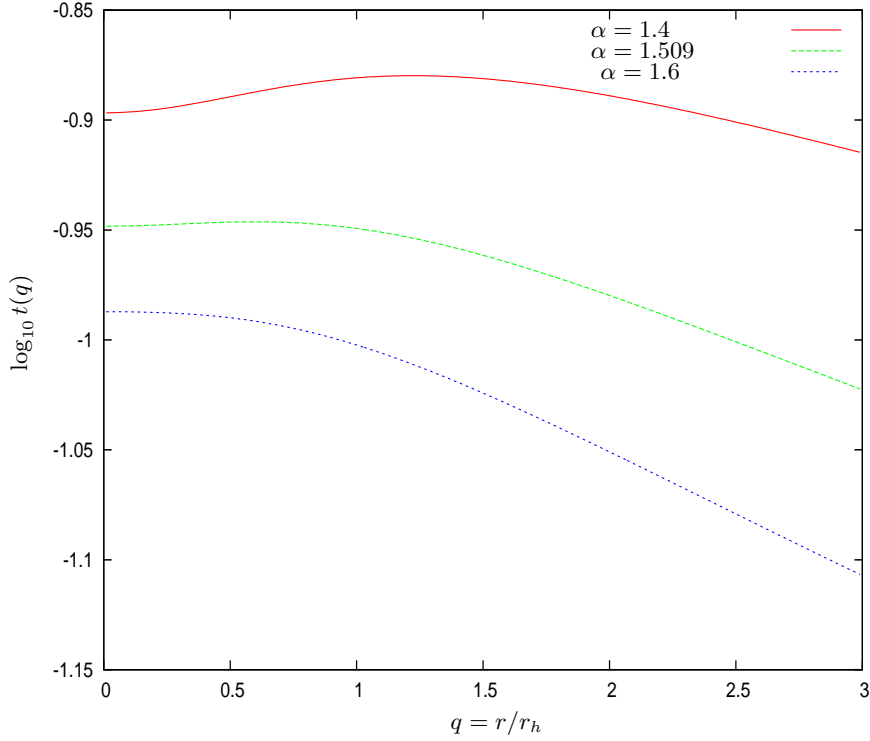}
\end{turn}
\caption{The ordinary logarithm of the normalized ratio
$ \frac1{4 \; \pi \; G \; \Sigma_0 \; r_h} \; \frac{P(r)}{\rho(r)} =
T(q)/[4 \; \pi \; m \; G \; \Sigma_0 \; r_h] = \frac{\Pi(q)}{4 \; \pi \; F(q)} = t(q) $ (or dimensionless temperature)
vs. $ q $ for three values of the exponent $ \alpha , \; 1.4 < \alpha \leq 1.6 $
in the density profile eq.(\ref{Falfa}). We see a variable temperature $ T(r) $.
Moreover, inside the halo radius $ r/r_h = q < 1 $ for $ \alpha = 1.509 $
the temperature is approximately constant indicating approximate thermalization.}
\label{psobrerho}
\end{figure}

We compute now the DM equation of state from the pressure eq.(\ref{vepre})-(\ref{defPI})
and the density profile eq.(\ref{Falfa}). We obtain from eqs.(\ref{pres}) and (\ref{vepre})
\be
\frac{P(r)}{\rho(r)} = \frac13 \; v^2(r) = G \; \Sigma_0 \; r_h \; \frac{\Pi(q)}{F(q)} \quad .
\ee
We plot in fig. \ref{psobrerho} the normalized ratio $ P(r)/\rho(r) $:
$$
\frac1{G \; \Sigma_0 \; r_h} \; \frac{P(r)}{\rho(r)}  \; ,
$$
versus the distance $ q $ for three relevant values of $ \alpha $.

{\vskip 0.2cm} 

We see that for $ \alpha \sim 1.5 $ inside the halo radius $ r < r_h $ ($ q < 1 $) 
the normalized ratio $ P(r)/\rho(r) $ and thus the velocity dispersion $ v^2(r)/3 $ turn to be 
{\bf approximately constant} (independent of $ r $). This implies that the dark matter obeys 
locally, at each point, an ideal gas equation of state
\be\label{Tq}
P(r) = \frac{T(r)}{m} \; \rho(r)  \quad , \quad T(r) \equiv 4 \; \pi \; m \; G \; \Sigma_0 \; r_h \; t(q) \quad , \quad
t(q) \equiv \frac{\Pi(q)}{4 \; \pi \; F(q)} \; .
\ee
The local temperature $ T(r) $ is related to the dispersion velocity by the usual relation
\be\label{Tv2}
 T(r) = \frac13 \; m \;  v^2(r) \quad .
\ee
The temperature $ T(r) $ is {\bf approximately constant} $ \simeq T_0 $ inside the halo radius for $ \alpha \sim 1.5 $
as shown by figs. \ref{psobrerho} and \ref{tef}. 
More precisely, for $ \alpha = 1.509 , \; T(q) $ is the closest to a constant inside the halo $ 0 < q < 1 $:
\be\label{valorT}
t(q) \simeq t_0 \simeq 0.113  \quad , \quad 0 < q < 1 \quad , \quad \alpha = 1.509 \; .
\ee
Namely, for $ \alpha = 1.509 $ dark matter in the galaxy halo is closest to thermal equilibrium for the density
profile eq.(\ref{Falfa}).

{\vskip 0.2cm} 

The density profile eq.(\ref{Falfa}) for $ \alpha = 1.5 $ (modified Hubble model)
provides a good approximation to the isothermal sphere density profile in the region $ 0 < q < 4 $ 
[sec. 4.3.3.b in ref. \cite{bin}].
This is a further support to the approximate halo thermalization put forward in the present paper.

{\vskip 0.2cm} 

We see that inside the halo, the value of the dimensionless temperature $ t(q) $ 
eq.(\ref{valorT}) obtained here from the equation of state, as well as
from the distribution function in eq.(\ref{fiteo}) only differ by 2 \%.

{\vskip 0.2cm} 

We see from eq.(\ref{T0K}) that the galaxy temperature $ T_0 $ grows {\bf linearly} with the halo radius.
$ T_0 $ can be expressed in terms of the halo mass using eq.(\ref{mscal})
with the coefficient $ d_\alpha $ corresponding to the case closest to thermal equilibrium, $ \alpha = 1.509, \;
d_{1.509} = 2.18188 $, with the result:
\be\label{scalTq}
r_h =  61.801 \; \sqrt{\frac{M_h}{10^6 \; M_\odot} \frac{120\; M_\odot}{\Sigma_0 \;  {\rm pc}^2}} 
\;\; \; {\rm pc} \quad , \quad T(q) = 103.5 \; t(q) \; \frac{m}{2 \; {\rm keV}} \; 
\sqrt{\frac{\Sigma_0 \; {\rm pc}^2}{120 \; M_\odot} \frac{M_h}{10^6 \; M_\odot}} \; \; {\rm mK} \; .
\ee
Hence, the galaxy temperature $ T_0 $ grows as the square root of the halo mass.

\medskip

For $ q \gtrsim 1 $, the local temperature decreases. 
In fig. \ref{tef} we plot $ t(q) $ vs. $ \log_{10} q $ for $ \alpha = 1.509 $. 
We see from fig. \ref{tef} that the local temperature  $ t(q) $
decreases slowly with $ q $. For instance,  $ t(q=10) \simeq \frac12 \; t(q<1) $
and $ t(q=100) \simeq 0.095 \; t(q<1) $. In other words, the DM is not in global thermal
equilibrium for $ q > 1 $ but there is a {\bf local}
thermal equilibrium with a local normalized temperature $ t(q) $
that decreases slowly when one gets farther from the center.

{\vskip 0.1cm} 

A local temperature for $ q \gtrsim 1 $ follows by fitting the distribution function 
$ {\cal D}(q) $ to a Boltzmann-like form for $ q \gtrsim 1 $ as in eq.(\ref{bolt}):
a local decreasing temperature $ t_0 $ with a similar behaviour to eq.(\ref{Tq}) is obtained,
showing that the concept of a local $q$-dependent temperature,
slowly varying with the coordinates is perfectly consistent in the region $ r \gtrsim r_h $
of the DM galaxy halo.

\subsection{Polytropic behaviour of the equation of state}\label{polyes}

\begin{figure}
\begin{turn}{-90}
\psfrag{"SLpLrho1509.dat"}{$ \alpha = 1.509 $}
\psfrag{"SLpLrhoplum.dat"}{$ \alpha = 2.5 $}
\psfrag{F(q)}{$ \log_{10} F(q) $}
\psfrag{Pi(q)}{$ \log_{10} \Pi(q) $}
\includegraphics[height=12.cm,width=10.cm]{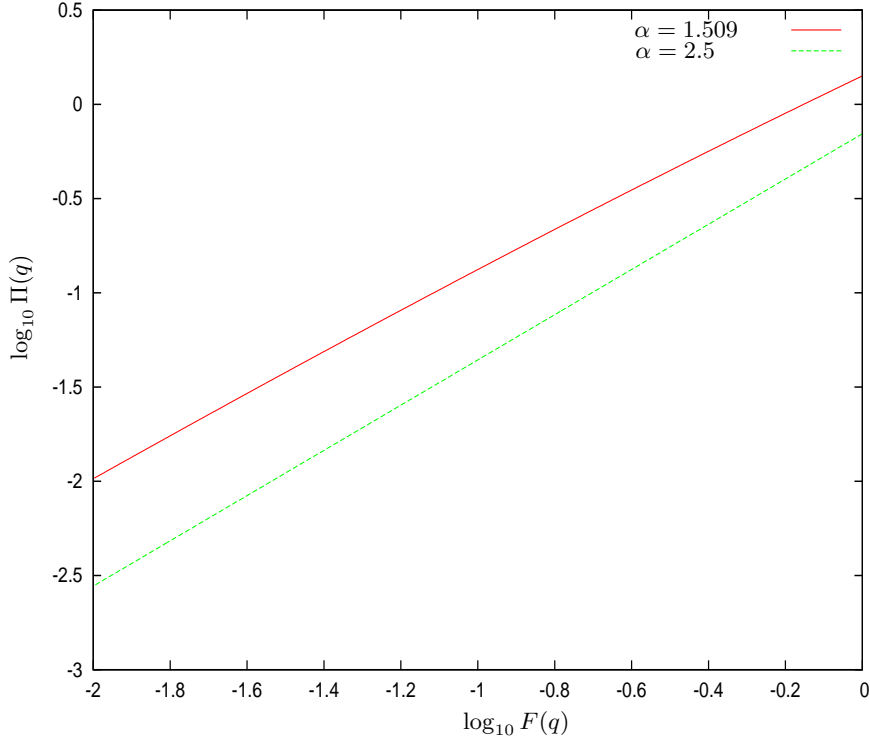}
\end{turn}
\caption{The ordinary logarithm of the normalized pressure $ \Pi(q) $ eq.(\ref{vepre})
vs. the ordinary logarithm of the normalized density $ F(q) $ eq.(\ref{denF})
for the dark matter cored profile $ \alpha = 1.509 $ and for the Plummer 
(stellar globular cluster) profile ($ \alpha = 2.5 $). The straight lines obtained
in this log-log plot correspond to the polytropic equations of state in 
 eq.(\ref{pipo}). }
\label{poly}
\end{figure}

We plot in fig. \ref{poly} the ordinary logarithm of the normalized pressure
$$
\Pi(q) = \frac{P(r)}{G \; \Sigma_0^2} \; ,
$$
as a function of the ordinary logarithm of the normalized density
$$
F(q) = \frac{\rho(r)}{\rho_0} \; ,
$$
where we used eqs.(\ref{denF}) and (\ref{vepre}).

{\vskip 0.1cm} 

We see from fig. \ref{poly} that in very good approximation (better than the percent) 
the pressure behaves as a power of the density:

\be\label{pipo}
P = A \; \rho^\gamma \quad , \quad \Pi = a \; F^\gamma \quad , \quad A = G \; r_h^2 \; \rho_0^{2-\gamma} \; a
\; ,
\ee
where $ A , \; a $ and $ \gamma $ are constants. 

{\vskip 0.1cm} 

We display in Table \ref{polyt} the values of $ \gamma $ and $ a $ for the DM halo 

profile $ \alpha = 1.509 $ and for the Plummer (stellar globular cluster) profile  $ \alpha = 2.5 $.
$ \gamma $ and $ a $ are obtained by a least-square fit to the numerical values of $ \Pi(q) $ and $ F(q) $
for $ \alpha = 1.509 $ in the relevant region $ 0 < r < R_{virial} $ 
 and from eq.(\ref{poly25}) in the Plummer case.
The obtained polytropic indexes are 
$$ 
\gamma_{halo} = 1.05  \quad {\rm and}  \quad \gamma_{Plummer} = 1.2 \; .
$$

In the Plummer case of stellar globular clusters ($ \alpha = 2.5 $)
the polytropic index $ \gamma = 1.2 $ is an exact result
valid for all density values eq.(\ref{poly25}) .

\begin{table}
\begin{tabular}{|c|c|c|} \hline  
\quad & & \;  \quad  
\\
\quad  $ \alpha $ \quad  & \quad  $ \gamma $ \quad   & \quad  $ a $ \;  \quad  
\\ 
\quad & &  \; \quad  
\\ \hline  
\quad  1.509 \quad   & \quad  1.05 \quad  & \quad  1.46 \;  \quad  
\\ \hline  
\quad  2.5 \quad  &\quad  1.2 \quad  & \quad  0.698 \;  \quad  
\\ \hline  
\end{tabular}
\caption{The polytropic index $ \gamma $ and the amplitude $ a $ in the polytropic equation
of state for the DM cored profile $ \alpha = 1.509 $ and
for stellar globular clusters described by the Plummer profile.}
\label{polyt}
\end{table}

\section{Galaxy Halos turn to be in approximate Local Thermal Equilibrium}\label{eqtloc}

Once the galaxy is formed, it is in a virialized stationary state
of size $ R_{virial} \; : \; 0 \leq r \leq R_{virial} $.

{\vskip 0.1cm} 

The virialized region is in an approximate {\bf local} thermal equilibrium situation, 
namely at each point $ r $,
the equation of state is that of an ideal self-gravitating gas with a local velocity
dispersion $ v^2(r) = 3 \; T(r) / m $ which slowly varies with $ r $. The local
temperature $ T(r) $ slowly varies with $ r $ too. 

{\vskip 0.1cm} 

Inside the halo radius $ r \lesssim r_h ,$
the magnitudes $ v^2(r) $ and $ T(r) $ vary with $ r $
as shown in fig. \ref{psobrerho}. We see a variable temperature $ T(r) $.
Moreover, inside the halo radius $ r/r_h = q < 1 $ for $ \alpha = 1.509 $
the temperature  $ T(r) $ is approximately constant indicating an approximate thermal equilibrium situation.

{\vskip 0.1cm} 

It is instructive to look to the analytic expression of $ t(q) $ for $ \alpha = \frac32 $
that follows from eqs.(\ref{pi15}) and (\ref{Tq}):
\be\label{tq15}
t(q) = \frac12  \sqrt{1+q^2} + (1+ 2 \; q^2) \; (1 + q^2)
\; \frac{{\rm Arg \, Sinh}(q)}{q}
- (1+q^2)^\frac32 \; \ln[4(1+q^2)] \quad , \quad  \alpha = \frac32 \; ,
\ee
Expanding this function in powers of $ q^2 $ yields
\bea
&& t(q) = \frac32 - 2 \; \ln2 + 3 \; \left(\frac{25}{36} - \ln2 \right) \; q^2
- \frac34 \; \left(\ln2 - \frac{41}{60} \right) \; q^4 + \frac18 \left(\ln2 - \frac{269}{420} \right) \; q^6
+ {\cal O}\left(q^8\right) = \cr \cr\cr
&& = 0.113706 \left[1 + 0.0342269 \; q^2 - 0.0647319 \; q^4 + 0.0579028 \; q^6
+ {\cal O}\left(q^8\right)\right]
\eea 
We see that this series which converges for $ |q| < 1 $, has small and alternating sign coefficients.
This is an explanation why $ t(q) $ is practically constant for $ \alpha \simeq \frac32 $
and $ q \lesssim 1.5 $ as shown in fig. \ref{psobrerho}.

{\vskip 0.2cm} 

\begin{figure}
\begin{turn}{-90}
\psfrag{"Npsrhoq.dat"}{$ t(q) $ vs. $ \log_{10} q $ for $ \alpha = 1.509 $}
\psfrag{"Ntcalfa5.dat"}{$ t_c(q) $ vs. $ \log_{10} q $ for $ \alpha = 1.509 $}
\psfrag{logq}{$ \log_{10} q $}
\psfrag{t(q)}{$ t(q) $ and $ t_c(q) $}
\includegraphics[height=12.cm,width=10.cm]{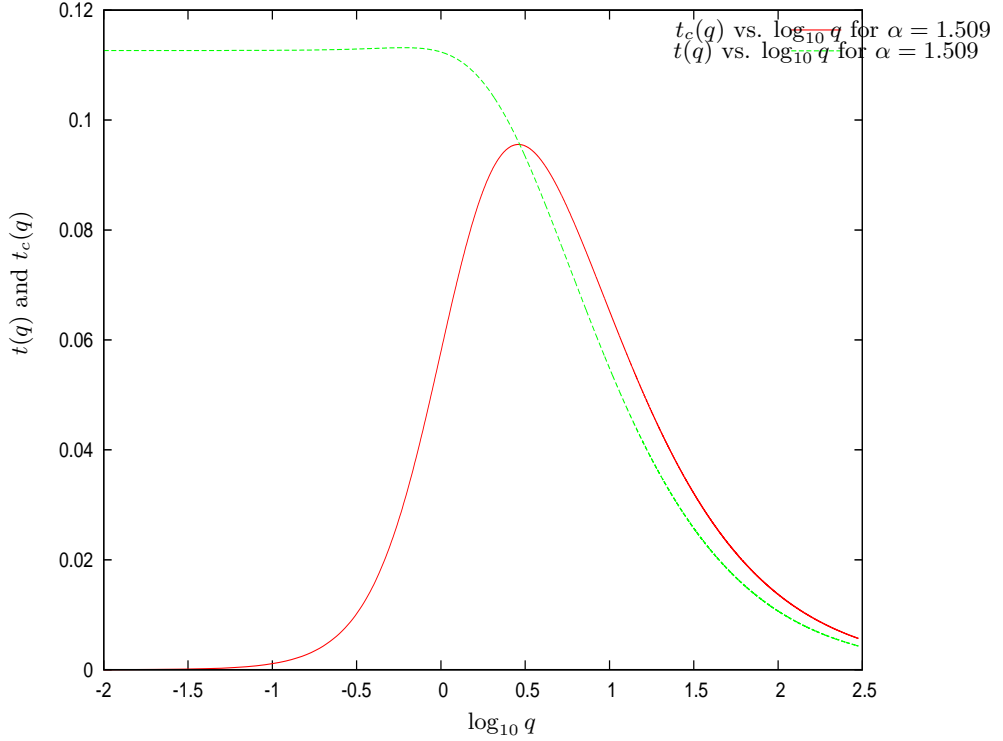}
\end{turn}
\caption{The normalized temperature $ t(q) $ and the circular temperature  $ t_c(q) $
vs. $ \log_{10} q $ for $ \alpha = 1.509 $.
It clearly displays approximate thermal equilibrium inside the halo radius $ q < 1 $.
For $ q \gtrsim 1 $ it shows local thermal equilibrium with a normalized temperature
$ t(q) $ that decreases slowly with $ q $ following the decreasing circular temperature $ t_c(q) $.}
\label{tef}
\end{figure}

Beyond the halo radius, in the region $ r_h \lesssim r \leq R_{virial} , \; v^2(r) $ and $ T(r) $
slowly decrease with $ r $ as shown in fig. \ref{tef}.

{\vskip 0.3cm} 

It is illuminating to consider here the circular velocity 
\be 
v^2_c(r) \equiv \frac{G \; M(r)}{r} \; ,
\ee
and in analogy with eq.(\ref{Tv2}), we associate to $ v^2_c(r) $
the circular temperature 
\be
T_c(r) \equiv \frac13 \; m \;  v_c^2(r) \; .
\ee
From eqs.(\ref{masar}) and (\ref{Tq}) we have respectively,
$$
v^2_c(r) = 4 \; \pi \; G \; \rho_0 \; r_h^2 \; \frac{w_\alpha(q)}{q} \quad , 
$$
\be\label{deftc}
T_c(r)= 4 \; \pi \; m \; G \; \rho_0 \; r_h^2 \; t_c(q) 
\quad , \quad t_c(q) \equiv \frac13 \; \frac{w_\alpha(q)}{q} \; .
\ee

We plot in fig. \ref{tef} the normalized temperature $ t(q) $ and the circular temperature  $ t_c(q) $
vs. $ \log_{10} q $ for $ \alpha = 1.509 $. We clearly see approximate thermal equilibrium 
inside the halo radius $ r < r_h, \; (q < 1) $.

{\vskip 0.1cm} 

Outside the halo $ r \gtrsim  r_h, \; (q \gtrsim 1) $ the local temperatures $ t(q) $ and $ t_c(q) $
decrease slowly with the distance to the center $ r $. Moreover, we see in fig. \ref{tef} 
that the local temperature $ t(q) $
{\bf follows} the decrease of the circular temperature  $ t_c(q) $ in this region $ q \gtrsim 1 $.
Around $ q = 10 $ we find $ t(q) \simeq 0.8 \; t_c(q) $

{\vskip 0.1cm} 

More precisely, we can compare the large distance asymptotic behaviour
of the local temperature $ t(q) $ for $ \alpha = 3/2 $ given by eq.(\ref{tq15}) 
with that of the circular temperature $ t_c(q) $ that follows from eqs.(\ref{w15}) and (\ref{deftc}):
\bea\label{tas}
&&t(q) \buildrel_{q \gg 1}\over= \frac1{4 \; q} \left[ \ln\left(2 \; q\right) - \frac34
\right] \; \left[1 + {\cal O}\left(\frac1{q^2}\right) \right]  \; , \cr \cr \cr 
&& t_c(q) \buildrel_{q \gg 1}\over= \frac1{3 \, q} \; \left[
\ln\left(2 \; q\right) - 1 \right] 
\; \left[1 + {\cal O}\left(\frac1{q^2}\right) \right]  \quad , 
\eea
We see from eq.(\ref{tas}) that the temperature $ t(q) $ and the circular temperature  $ t_c(q) $
exhibit the same analytic structure for $ q \gtrsim 5 $. 
Asymptotically, $ t_c(q) $ is larger than $ t(q) $ by a factor
$ 4/3 $:
\be\label{tas2}
\frac{t_c(q)}{t(q)} \buildrel_{q \gg 1}\over= \frac43 \; \left[1 - \frac1{4 \; \ln\left(2 \; q\right) -3}  
+ {\cal O}\left(\frac1{q^2}\right) \right]  \quad , 
\ee
As shown by fig. \ref{tef}, $ t_c(q) $ is larger than $ t(q) $ by 
a factor going approximately from unity at $ q = 2.93 $ to 1.3 at $ q = 160 $ and reaching
the limiting value $ 4/3 $ at $ q = \infty $ as follows from eq.(\ref{tas2}).

{\vskip 0.1cm} 

Moreover, it must be noticed that the temperature $ T_0 $  eq.(\ref{valorT}) obtained from the velocity dispersion
eq.(\ref{Tv2}) coincides better than 2 \% with the temperature obtained from the slope of the DF
$ \ln {\cal D}(-\varepsilon) $ eq.(\ref{Dq}) through eqs.(\ref{bolt})-(\ref{tcero}).
This agreement shows the consistency of our approach.

{\vskip 0.1cm} 

In addition, the local temperature $ T(r) $ for $ r \gtrsim r_h $ derived from the slope of 
$ \ln {\cal D}(-\varepsilon) $ and shown in fig.  \ref{lnD} and $ T(r) $ derived from the equation of state
as shown in figs. \ref{psobrerho} and  \ref{tef} decrease in a remarkably similar way.

{\vskip 0.2cm}

It is instructive to rewrite  eq.(\ref{tcero}) as
\be \label{tcero2}
\frac{T_0}{m} = 4 \; \pi \; G \; \Sigma_0 \; r_h \; t_0  \quad , 
\ee
we see that the value of $ T_0/m $ follows from galaxy observations and the value of $ t_0 \simeq 0.11 $
from eqs.(\ref{fiteo}) and (\ref{valorT}).  $ t_0 $ is an universal galaxy independent number.
Observations from dilute galaxies with halo masses $ M_h > 10^6 \; M_\odot $ cannot determine
$ T_0 $ and $ m $ separately, only the ratio $ T_0/m $ is obtained.
The physical reason why once we
know $ r_h $ and $ \Sigma_0 $ we can compute $ T_0/m $ is that we use the velocity
dispersion information.

{\vskip 0.2cm}

Physically, these phenomena are clearly understood because in the inner halo region
$ r \lesssim r_h $, the density is higher than beyond the halo radius.
The gravitational interaction in the inner region is strong enough and
thermalizes the self-gravitating gas of DM particles while beyond the halo radius
the particles are too dilute to thermalize, namely, although they are virialized,
they had not enough time to accomplish thermalization. Notice that virialization always starts 
before than thermalization. 

{\vskip 0.1cm} 

In the process of thermalization there is an energy transfer flow of potential energy into kinetic energy.
Clearly, in the outside halo region $ r \gtrsim  3 \; r_h $ we find that the kinetic energy is lower
than in the inside the region $ r < r_h $ where thermalization is already achieved.
Therefore, the local temperature $ T(r) $ in the outside halo region $ r \gtrsim  3 \; r_h $
is lower than the temperature $ T_0 $ in the internal region $ r < r_h $ where thermalization is achieved.

{\vskip 0.1cm} 
 
This treatment applies to dilute large galaxies which are in a classical
physics regime for halo masses $ M_h > 10^6 \; M_\odot $.

{\vskip 0.1cm} 
 
For smaller (dwarf) galaxies there is not yet available information on density profiles
from observations in order to determine $ F(q) $. Knowing $ F(q) $ for dwarf galaxies
will allow to apply the methods developed here to find the phase--space distribution
function $ f(E) $ and the velocity dispersion $ v^2(r) $ for dwarf galaxies.

{\vskip 0.1cm} 

Notice that the Thomas-Fermi approach to galaxy structure
applies to {\bf all types of galaxies} and allows to determine
theoretically all physical magnitudes for them \cite{newas}-\cite{astro}.
In the classical regime, for halo masses $ M_h > 10^6 \; M_\odot $, the
galaxy equation of state computed in the Thomas-Fermi approach yields
the same results as found here from the empirical $ \alpha $-profiles for $ \alpha \simeq 1.5 $
and the Eddington equation, namely the Boltzmann self-gravitating ideal gas equation
of state, which shows the robustness of these results.

{\vskip 0.1cm} 

This paper provides an unified framework in which galaxy structure is
obtained from the collisionless self-gravitating DM gas of particles in analogous way as globular
clusters structure is obtained from a gas of stars.


\acknowledgments

We thank Paolo Salucci for useful discussions.

\section{Appendix}

We compute in this Appendix the integral eq.(\ref{intap}).

We insert the integral representaion for $ \Psi(-\nu') $ given by eq.(\ref{psinu})
$$
J(\nu) \equiv \int_{\nu(\infty)}^{\nu} d\nu' \; \left(\nu-\nu' \right)^\frac32 \;  \Psi(-\nu')=
\sqrt2 \; \pi \; \frac{\rho_0}{m^\frac52 \; T_0^\frac32} \;
\int_{\nu(\infty)}^{\nu} d\nu' \; \left(\nu-\nu' \right)^\frac32 \; 
\int_{\nu(\infty)}^{\nu'} \frac{d\nu''}{\sqrt{\nu'-\nu''}} \; \frac{d^2 F}{d\nu''^2}\; .
$$
Interchanging the integration over $ \nu' $ and $ \nu'' $ yields
$$
J(\nu) = \sqrt2 \; \pi \; \frac{\rho_0}{m^\frac52 \; T_0^\frac32} \; \int_{\nu(\infty)}^{\nu} \; d\nu''
\; \frac{d^2 F}{d\nu''^2} \int_{\nu''}^{\nu} \; \frac{\left(\nu-\nu' \right)^\frac32}{\sqrt{\nu'-\nu''}} \; d\nu' \; .
$$
The integral over $ \nu' $ can be performed explicitly with the result
$$
\int_{\nu''}^{\nu} \; d\nu' \; \frac{\left(\nu-\nu' \right)^\frac32}{\sqrt{\nu'-\nu''}} = \frac38
\; \pi \; \left(\nu-\nu'' \right)^2 \; ,
$$
and therefore,
$$
J(\nu) =\frac{3 \, \pi^2}{4 \, \sqrt2} \; \frac{\rho_0}{m^\frac52 \; T_0^\frac32} \; 
\int_{\nu(\infty)}^\nu d\nu' \; \left(\nu-\nu'\right)^2 \; \frac{d^2 F}{d\nu'^2} \; .
$$
as in eq.(\ref{intap}).

\end{document}